\documentclass[doublespacing]{elsart}
\usepackage{icarus}

\usepackage{natbib}

\bibpunct{(}{)}{;}{a}{,}{,}

\usepackage{graphicx}

\newcommand{\htwo}{\mbox{H${}_2$}}
\newcommand{\chfour}{\mbox{CH${}_4$}}

\newcommand{\ctwohtwo}{\mbox{C${}_2$H${}_2$}}
\newcommand{\ctwohfour}{\mbox{C${}_2$H${}_4$}}
\newcommand{\ctwohsix}{\mbox{C${}_2$H${}_6$}}
\newcommand{\cfourhtwo}{\mbox{C${}_4$H${}_2$}}

\newcommand{\degr}{$^\circ$}

\newcommand{\dw}{$\Delta$W}
\newcommand{\dl}{$\Delta$L}

\def\scinot#1.{\hbox{$\,$ $\times$ $10^{#1}$}}

\def\deg{\ifmmode^\circ\else$\null^\circ$\fi}
\def\spose#1{\hbox to 0pt{#1\hss}}
\def\lta{\mathrel{\spose{\lower 3pt\hbox{$\mathchar "218$}}\raise 2.0pt\hbox{$\mathchar "13C$}}}
\def\gta{\mathrel{\spose{\lower 3pt\hbox{$\mathchar "218$}}\raise 2.0pt\hbox{$\mathchar "13E$}}}

\begin{document}

\begin{frontmatter}



\title{Saturn's upper atmosphere during the Voyager era: Reanalysis and modeling of the UVS occultations}


\author{Ronald J. Vervack, Jr.}

\address{The Johns Hopkins University Applied Physics Laboratory \\
         11100 Johns Hopkins Road \\
         Laurel, MD 20723-6099 (U.S.A.)}

\author{Julianne I. Moses}

\address{Space Science Institute \\
         4750 Walnut Street \\
         Suite 205 \\
         Boulder, CO  80301 (U.S.A.)}

%% This copyright statement isn't required at any stage by the Icarus
%% Editorial Office or Elsevier.  However, until you sign over the
%% copyright to Elsevier prior to publication (or negotiate with them
%% about copyright), you own the copyright to anything you create.
%% Just to keep things unambiguous, always include a copyright statement
%% or explicitly dedicate your work to the public domain.
\begin{center}
\scriptsize
Copyright \copyright\ 2014 Ronald J. Vervack, Jr.
\end{center}

%% ----- ELSEVIER STUFF -----
%% The commands below up to the \end{frontmatter} are commented out
%% so that we can do some Icarus-required formatting on the second and
%% third pages that is not required later on by Elsevier.  So when
%% your paper gets accepted, and you are ready to start dealing with
%% Elsevier, copy your abstract and keywords up here, uncomment these
%% lines, and comment out the ICARUS STUFF below.
%% 
%% Alternately, you might just want to move these abstract, keyword,
%% and end frontmatter commands down, and comment out the ICARUS STUFF
%% commands.  It doesn't matter.

% \begin{abstract}
% % Text of abstract
% 
% \end{abstract}
% 
% \begin{keyword}
% % keywords here, in the form: keyword \sep keyword
% 
% 
% % PACS codes here, in the form: \PACS code \sep code
% 
% \end{keyword}

%% ----- END ELSEVIER STUFF -----

\end{frontmatter}

%% ----- ICARUS STUFF -----
%% Some formatting on the first, second, and third pages are required
%% by the Icarus Editorial Office that are not required by Elsevier.
%% This section contains those things.  When you are ready to transition
%% to ``Elsevier'' mode, copy your abstract and keywords up into
%% the ELSEVIER STUFF section, and then you can just delete everything
%% in this section.

%% We need to list the number of manuscript pages, figures, and tables. 
%%
%% Rather than manually count these things out, we'll use a little
%% trick here from Paul.  All you have to do is place three \label{}
%% tags on the last page, the last table, and the last figure, that
%% way these values are automatically updated (as long as you remember
%% to move the lasttable and lastfig labels when you add or remove
%% tables and figures).

\begin{flushleft}
\vspace{1cm}
Number of pages: \pageref{lastpage} \\
Number of tables: \ref{lasttable}\\
Number of figures: \ref{lastfig}\\
\end{flushleft}

%% Don't worry about finding the various last* tags and deleting them
%% when you go to ``Elsevier'' mode if you don't want to, they should be
%% silently ignored.

%% The second page should indicate a proposed running head of not more 
%% than 55 characters, and the name and address to which editorial 
%% correspondence and proofs should be directed.  The pagetwo 
%% environment that icarus.sty provides will make page two for you,
%% just give the running head as an argument to the environment, and
%% then your correspondence address inside.
\begin{pagetwo}{Saturn's upper atmosphere during the Voyager era}
%                        1         2         3         4         5
%               1234567890123456789012345678901234567890123456789012345

Ronald J. Vervack, Jr. \\
The Johns Hopkins University \\
Applied Physics Laboratory \\
11100 Johns Hopkins Road \\
Laurel, MD 20723-6099, USA. \\
\\
Email: Ron.Vervack@jhuapl.edu \\
Phone: (443) 778-8221 \\
Fax: (443) 778-1641

\end{pagetwo}

\begin{abstract}
The Voyager 1 and 2 Ultraviolet Spectrometer (UVS) solar and stellar
occultation dataset represents one of the primary, pre-Cassini sources
of information that we have on the neutral upper atmosphere of Saturn.
Despite its importance, however, the full set of occultations has
never received a consistent, nor complete, analysis, and the results
derived from the initial analyses over thirty years ago left questions
about the temperature and density profiles unanswered.  We have
reanalyzed all six of the UVS occultations (three solar and three
stellar) to provide an up-to-date, pre-Cassini view of Saturn's upper
atmosphere. From the Voyager UVS data, we have determined vertical
profiles for \htwo, H, \chfour, \ctwohtwo, \ctwohfour, and \ctwohsix,
as well as temperature.  Our analysis also provides explanations for
the two different thermospheric temperatures derived in earlier
analyses (400-450~K versus 800~K) and for the unusual shape of the
total density profile noted by \citet{Hubbard97}.  Aside from
inverting the occultation data to retrieve densities and temperatures,
we have investigated the atmospheric structure through a series of
photochemical models to infer the strength of atmospheric mixing and
other physical and chemical properties of Saturn's mesopause region
during the Voyager flybys.  We find that the data exhibit considerable
variability in the vertical profiles for methane, suggesting
variations in vertical winds or the eddy diffusion coefficient as a
function of latitude and/or time in Saturn's upper atmosphere.  The
results of our reanalysis will provide a useful baseline for
interpreting new data from Cassini, particularly in the context of
change over the past three decades.

\end{abstract}

% %% Keywords should appear after the abstract. 
\begin{keyword}
Saturn, Atmosphere\sep Occultations\sep Aeronomy\sep
Atmospheres, Structure\sep Ultraviolet Observations
\end{keyword}

%% ----- END ICARUS STUFF -----

%main text

%\figcapson
%\printfigures

\section{Introduction}
\label{sec:intro}

\begin{figure}[p]
    \includegraphics[width=6.5in]{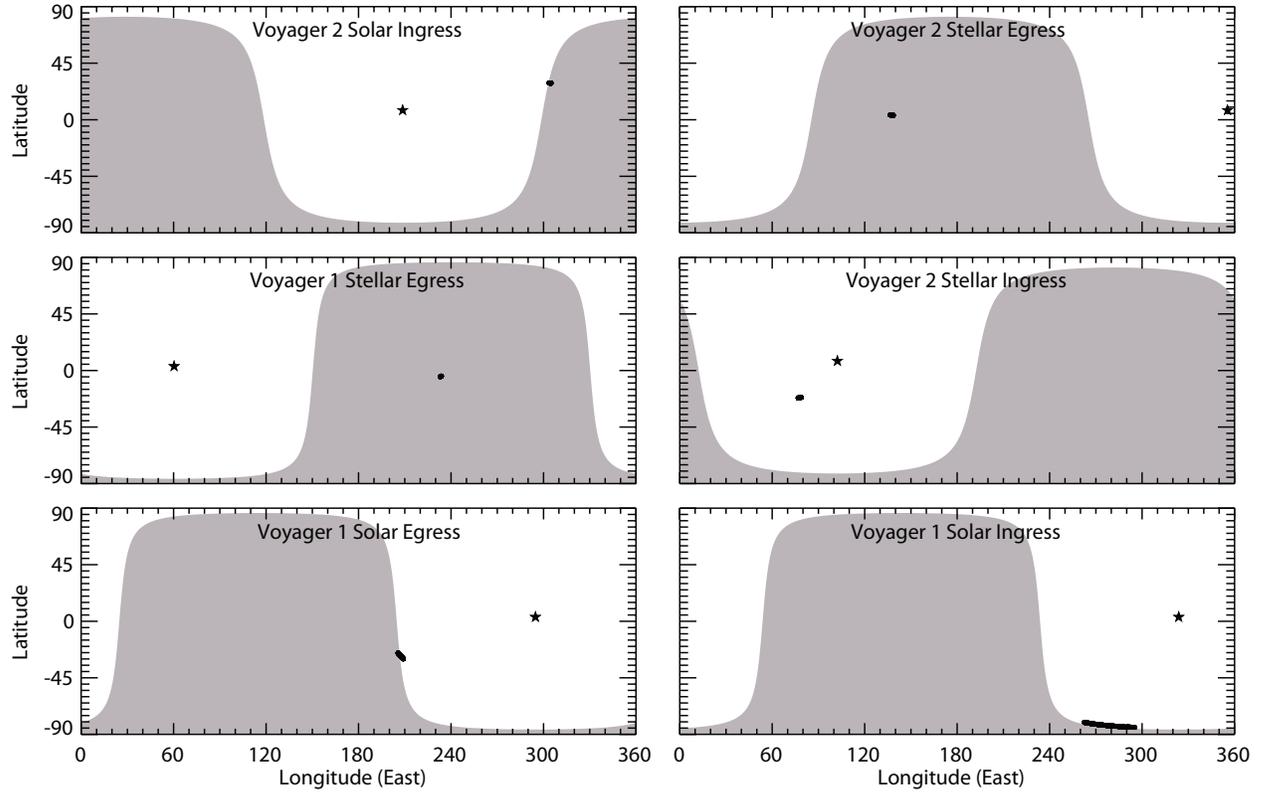}
    \caption{Geometry of the six Voyager~1 and~2 UVS Saturn
    occultations.  The planetocentric latitude and longitude of the
    tangent point (the point of closest approach) of each observed
    spectrum is plotted as a circle, resulting in a line spanning the
    range of latitude and longitude probed during a given occultation.
    The subsolar point is indicated by a star in each panel, and the
    shaded region indicates the night side of Saturn.  For most
    occultations, the range probed is small, but owing to the rapid
    trajectory changes during the Voyager~1 solar occultations, the
    ranges are larger.  Of note is the Voyager~2 stellar ingress,
    which is the only occultation that occurred completely on the day
    side of Saturn.}
    \label{fig:geometry}
\end{figure}

The Voyager 1 and 2 spacecraft encountered the Saturn system during
November 1980 and August 1981, respectively.  During these encounters,
the Ultraviolet Spectrometer (UVS) on the two spacecraft performed six
occultation experiments at Saturn.  Voyager 1 carried out ingress and
egress solar occultations and an egress stellar occultation of the
star $\iota$ Herculis, while Voyager 2 performed an ingress solar
occultation and both an ingress and egress occultation of the star
$\delta$ Scorpii.  The geometry of these occultations is summarized in
Fig.~\ref{fig:geometry} and Table~\ref{tab:geometry}.

\clearpage

\begin{table}[h]
  \caption{Summary of Geometry for the Voyager 1 and 2 UVS Saturn Occultations}
  \begin{tabular}{lccccc}
    \hline
                & UT         & Planetocentric  & Sub-Solar  & \\
    \cline{2-4}
                & Date       & Latitude Range  & Latitude   & Local Time$^{\rm a}$ \\
    Occultation & Time Range & Longitude Range & Longitude  & Range \\
    \hline
    Voyager 2 solar ingress   & 26 Aug 1981          & $29\deg$ to $29\deg$   & $8.1\deg$ & 8.23 to 8.26 \\
                              & 04:05:49 to 04:07:57 & $304\deg$ to $305\deg$ & $209\deg$ & \\

    Voyager 2 stellar egress  & 25 Aug 1981          & $3.5\deg$ to $3.9\deg$ & $8.1\deg$ & 9.60 to 9.66 \\
                              & 23:44:17 to 23:48:21 & $137\deg$ to $139\deg$ & $355\deg$ & \\

    Voyager 1 stellar egress  & 12 Nov 1980          & $-5.0\deg$ to $-4.5\deg$ & $3.9\deg$ & 10.55 to 10.57 \\
                              & 22:58:09 to 23:00:14 & $233\deg$ to $233\deg$   & $60\deg$  & \\

    Voyager 2 stellar ingress & 25 Aug 1981          & $-21.4\deg$ to $-21.9\deg$ & $8.1\deg$ & 4.62 to 4.68 \\ 
                              & 20:34:15 to 20:38:21 & $77\deg$ to $79\deg$       & $102\deg$ & \\ 

    Voyager 1 solar egress    & 13 Nov 1980          & $-30\deg$ to $-25\deg$ & $3.9\deg$ & 2.72 to 2.82 \\
                              & 02:39:34 to 02:44:46 & $206\deg$ to $209\deg$ & $295\deg$ & \\

    Voyager 1 solar ingress   & 13 Nov 1980          & $-84\deg$ to $-80\deg$ & $3.9\deg$ & 3.55 to 4.53 \\
                              & 01:47:57 to 01:53:21 & $262\deg$ to $295\deg$ & $324\deg$ & \\
    \hline
  \end{tabular}

  $^{\rm a}$ Local time is defined using a Saturn rotational period of 10.76
             hours (i.e., ``noon'' is 5.38 and ``midnight'' is 10.76).
  \label{tab:geometry}
\end{table}
\clearpage

The UVS occultation data consist of spectra taken before and during
atmospheric attenuation, from which the line-of-sight optical depth
can be determined.  From the wavelength and altitude variation of this
optical depth information, it is possible to infer both the densities
of the absorbing species and temperature information.  The early
analyses of the occultation data were reported by
\citet{Broadfoot81sat} and \citet{Sandel82sat}.  These analyses were
followed by the more in-depth work of \citet{Festou82} and
\citet{Smith83}.

Although the above analyses provided some information on the density
profiles of \htwo\ and \chfour, a disagreement in the value of the
thermospheric temperature has been the more profound result.  The
analyses of \citet{Broadfoot81sat} and \citet{Festou82} inferred a
temperature of 800--850~K, while those of \citet{Sandel82sat} and
\citet{Smith83} found a value between 400--450~K.  Currently, the
lower value is favored, particularly in light of the Cassini results
\citep{Shemansky12,Koskinen13}; however, the discrepancy has never
been satisfactorily explained and is puzzling given that it arises
from analysis of the same data.

Ground-based measurements conducted in the 1990's have added
additional concerns about previous analyses of the UVS Saturn
occultations.  The occultation of 28 Sgr by Saturn has been analyzed
by \citet{Hubbard97}, who find that the total density profile (and,
correspondingly, the temperature profile) in Saturn's mesosphere
inferred by \citet{Smith83} deviates significantly from the profiles
derived for Saturn's mesosphere from ground-based stellar occultations
(see \citeauthor{Hubbard97} Figure~13).  Because the
\citeauthor{Hubbard97} profiles merge well with the profile at lower
altitudes derived from the Voyager radioscience investigation, and
because the lowest 200~km of the UVS profile exhibits unusual
structure compared to the \citeauthor{Hubbard97} results, we suspect a
problem likely exists in the older analyses of the UVS data.

Illustrations of the potential significance of reanalyzing the Saturn
UVS occultation data are provided by the work of \citet{Yelle96} and
\citet{Vervack04}.  These authors reanalyzed the UVS occultations of
the star $\alpha$ Leonis by Jupiter and the Sun by Titan.  Their
results were notably different from those of previous analyses
\citep{Broadfoot81over,Festou81,Smith82}, particularly regarding the
temperatures in the upper atmosphere.  Although derived from the same
datasets, the improved processing and retrieval techniques utilized by
\citeauthor{Yelle96} and \citeauthor{Vervack04}, in combination with
other measurements \citep{Marten94,Hubbard95,Liu96} in the case of
\citeauthor{Yelle96}, resulted in significantly different temperature
structures in both the newer analyses.  The large temperature gradient
in Jupiter's lower thermosphere derived by \citeauthor{Yelle96} was
subsequently confirmed by the Galileo probe \citep{Seiff97}, while the
\citeauthor{Vervack04} results for Titan were confirmed by Cassini
INMS observations \citep{Waite04}.  Furthermore, the newer processing
and retrieval techniques have allowed density profiles for additional
species to be inferred from the data.

With the results from the Cassini Ultraviolet Imaging Spectrograph
(UVIS) stellar and solar occultation data now becoming available
\citep{Shemansky12,Gustin12,Koskinen13}, it would be valuable to
compare the state of the atmosphere in the current era with what was
observed thirty years ago.  The results may help further our
understanding of the source of the unexpectedly high temperatures on
all the giant planets, the details of the chemistry and dynamics in
thermosphere and middle atmosphere of Saturn, and the response of
Saturn's atmosphere to seasonal radiative forcing.

Deriving improved information on atmospheric structure and composition
in Saturn's upper atmosphere is important for many reasons.  The
mesopause region probed in the lower portion of the occultations,
i.e., the boundary between the middle atmosphere and thermosphere, is
a very interesting region from a physical and chemical standpoint.  At
deeper levels, atmospheric motions act to keep the lower atmosphere
well mixed, and the main stratospheric constituents --- hydrogen,
helium, and methane --- have mole fractions that are roughly constant
with altitude.  The ``eddy'' diffusion coefficient provides a means
for parameterizing the strength of this mixing in one-dimensional
models.  In the region probed by the UVS occultations, the mean free
path of the atmospheric molecules becomes large, and molecular
diffusion begins to dominate over eddy diffusion.  The pressure level
at which the molecular diffusion coefficient equals the eddy diffusion
coefficient is termed the ``homopause''.  Within several scale heights
of the homopause, the atmospheric composition varies dramatically with
height, as the concentration of each species becomes diffusively
controlled and begins to follow its own scale height.  Because most
atmospheric constituents are heavier than the background \htwo\ gas,
species concentrations drop off sharply near the homopause.

Coincident with the dramatic change in upper-stratospheric composition
is a dramatic change in thermal structure, as temperatures switch from
being low and nearly isothermal in the middle atmosphere to having a
large temperature gradient transitioning into the high-temperature
thermosphere.  The composition and thermal changes are related:
complex hydrocarbons like ethane and acetylene that are produced from
methane photochemistry help cool the stratosphere, and that cooling is
ineffective once the concentrations of these species drop off due to
molecular diffusion \citep[see][]{Yelle01}.  In addition, local
thermodynamic equilibrium breaks down in the low-density,
high-altitude regions, and the corresponding less effective cooling
through the ro-vibrational bands of the hydrocarbons may contribute to
the dramatic temperature increase \citep[e.g.,][]{Bezard97}.  Better
constraints on the location and properties of the temperature
structure in the homopause region would allow investigators to examine
the relative role of these two processes.  In addition, the location
of the homopause influences atmospheric chemistry through
pressure-dependent reactions \citep[e.g.,][]{Moses05etal}, and the UVS
occultations help pinpoint the homopause level and thus help constrain
the photochemical models.  The UVS occultations provide important
constraints for theoretical models of middle-atmospheric chemistry,
ionospheric chemistry, radiative transport, energy balance, and
atmospheric dynamics.

Given the unique insights into physical and chemical processes the
Voyager UVS datasets can provide, the improvements in analysis
techniques demonstrated by \citet{Yelle96} and \citet{Vervack04}
\citep[see also][]{Koskinen13}, and the problems noted by
\citet{Hubbard97}, there are a number of reasons to reanalyze the UVS
occultations at Saturn.  Perhaps the dominant reason, however, is that
the full set of six UVS occultations at Saturn has never received a
complete and consistent analysis owing to a variety of complicating
factors.  Therefore, we have revisted the entire Voyager UVS
occultation dataset to resolve the outstanding issues.  Using the
retrieved density and temperature information from our reanalysis, we
have also investigated the structure of Saturn's atmosphere through
photochemical modeling to provide a more complete picture of Saturn's
upper atmosphere during the Voyager~1 and~2 flybys and to set the
stage for comparisons of Voyager and Cassini observations.

\section{UVS Occultation Observations and Data Reduction}
\label{sec:datareduct}

The theoretical basis for absorptive occultations is the relatively simple
Lambert-Beer law,
\begin{equation}
  I(\lambda) = I_o(\lambda)\, e^{-\tau(\lambda)} {\rm ,} \label{eq:beer}
\end{equation}
where $I(\lambda)$ is the intensity of light at wavelength $\lambda$
after attenuation by some amount of absorbing material (in our case,
the atmosphere of Saturn), $I_o(\lambda)$ is the unattenuated
intensity, and $\tau(\lambda)$ is the optical depth of the absorbing
material (see \citet{Smith90} for a more detailed discussion of the
general theory of absorptive occultations).  In the case of each
Voyager UVS occultation, the UVS obtained a series of attenuated
spectra (referred to by $I$) as the line of sight to the source --- either the Sun or
a star --- moved through Saturn's atmosphere.  These spectra can be
divided by unattenuated reference spectra (referred to by $I_o$) acquired outside of
Saturn's atmosphere to construct a series of transmission spectra
(the ratio $I/I_o$) that are independent of the absolute UVS calibration.  The
variation of these transmission spectra with wavelength and altitude
can be analyzed to determine the abundance of UV absorbers in Saturn's
upper atmosphere.  However, these spectra are the convolution of the
transmission of Saturn's atmosphere with several instrumental and
observational effects related to the Voyager UVS.  These effects must
be removed from the observed spectra before analysis can begin.

\subsection{Description of the UVS}
\label{sec:uvsdescrip}

The Voyager~1 and~2 UVS instruments are compact, Wadsworth-mounted,
objective grating spectrometers covering slightly different wavelength
ranges: approximately 530--1700~\AA\ for Voyager~1 and 510--1680~\AA\
for Voyager 2.  They were designed to cover diagnostic wavelength
regions relevant to the major UV absorbers predicted or known to
constitute the upper atmospheres of the giant planets and some of
their satellites.  Detailed information on the design and operation of
the UVS may be found in \citet{Broadfoot77uvs},
\citet{Broadfoot77ccd}, \citet{Vervack97}, and \citet{Vervack04}.

Light enters the UVS through one of two ports and passes through a
mechanical collimator consisting of thirteen baffles before undergoing
a single normal incidence reflection from the concave diffraction
grating.  The airglow port is used to observe extended sources and
stars, whereas the occultation port is used to observe the Sun.  The
reflected light is then simultaneously focused and dispersed onto the
detector assembly, where all photons reaching the active area undergo
a three-stage process to convert the photon energy to detector counts.
Photons are first converted to charge using a semi-transparent
photocathode and two microchannel plates (MCP) in series.  The UVS can
operate in several different MCP gain states to enable it to observe
sources over a wide range of intensities.  The charge emitted from the
MCPs is collected by the detector, which is a self-scanned anode array
with 126 active elements spanning a wavelength interval of 9.26~\AA\
each.  Finally, the measured charge is converted to detector counts
via a low-resolution (3-bit) A-to-D converter.

Several effects related to the UVS design must be considered in the
analysis of UVS occultation spectra.  The use of a mechanical
collimator leads to only a fraction of the incoming light reaching the
grating as the source moves off-axis, and the reduction is described
by an instrumental ``slit function''.  The resolution of UVS spectra
varies with the source, being $\sim$30~\AA\ for extended (i.e.,
slit-filling) sources and $\sim$18~\AA\ for point sources.  The Sun is
too large to be a point source but not large enough to be an extended
source because it does not fill the slit.  The Sun is thus a
``finite'' source with a spectral resolution that falls between these
values.  Imperfections in the grating and the collimator result in the
scattering of some photons.  These ``scattered light'' photons are not
properly focused and result in radiation of wavelength~X contaminating
the image of the source for wavelength~Y and vice versa for all
wavelengths.  Finally, when the observed source moves off-axis, the
nominal wavelength range of the measured spectra shifts because of
changes in the incidence angle of the incoming light at the grating.
These last two effects are particularly important in interpreting the
wavelength-dependent absorption features in the transmission spectra.

\subsection{Spectral Processing}
\label{sec:processing}

Processing of UVS spectra prior to generating the $I/I_o$ spectra
requires the removal of several instrumental effects, including
anode-to-anode variations in detector sensitivity, the
charge-dependent spreading of the MCP-generated electron cloud, and
the presence of instrumentally scattered light.  For solar
occultations the UVS response is nonlinear owing to the high solar
intensities and must be corrected for the ratio of $I$ to $I_o$ to be
an accurate measure of the atmospheric transmission.  Additionally,
the even-numbered channels of both the Voyager~1 and~2 UVS detectors
suffer from a problem at high counting rates.  The cause is unknown
but it results in fewer counts being recorded than should be and the
problem can be particularly severe for solar observations.  Except for
scattered light, all these effects were removed through the use of a
detailed model of the UVS detector assembly.  Only a brief
summary of the processing is provided here; detailed descriptions of
the procedures and the detector assembly model may be found in
\citet{Vervack97} and \citet{Vervack04}.

\paragraph*{Detector assembly model.}
In order to relate the number of incoming photons at the grating to
the recorded counts on the detector, we have developed a mathematical
model of the entire detector assembly operation.  This model is based
on Monte Carlo methods that realistically simulate the passage of a
photon from the triggering event at the MCPs that initiates the
electron cascade, the spreading of the resulting electron cloud onto the
detector anodes, and the collection and conversion of those electrons
to detector counts.  The parameters of the model are based on a
rigorous study of both laboratory and in-flight measurements.  In
particular, the model utilizes laboratory measurements of the charge
spreading that have not been fully incorporated previously and
observations that have revealed changes in the MCP gains in-flight,
most likely a result of normal lifetime degradation of the MCPs.  The
latter effect is particularly severe in the case of the Voyager~1 UVS,
owing to the harsh radiation environment suffered during the Jupiter
encounter \citep[see][for a discussion]{Holberg82}.

For the Voyager 1 egress solar occultation, an intentional
mid-occultation change from the low (HVL~2) gain state to the high
(HVL~3) gain state made accurate knowledge of the detector gain
particularly important.  In the high gain state, solar spectra are
sufficiently bright to saturate the UVS at many wavelengths; thus, all
the reference $I_o$ spectra and roughly half of the $I$ spectra were
obtained in HVL~2.  Changing to HVL~3 midway through the occultation
after the brightest parts of the solar spectrum were attenuated
avoided saturation in all but a few instances at H Lyman $\alpha$ (none
in absorption or utilized reference spectra regions) and allowed the
occultation to probe deeper into Saturn's atmosphere.  To ensure
continuity in the transmission spectra with altitude, this change had
to be handled carefully.

Fortunately, two types of observations enabled the gains of the MCPs
to be inferred in-flight.  The UVS can operate in two counting modes:
pulse counting in which a single count is recorded when the charge
exceeds a single threshold and pulse height (also called pulse
integration) in which a series of counting thresholds allows counts to
range from 0 to 7.  Observations of the stars $\beta$ Centauri and
$\epsilon$ Persei in the two counting modes allowed us to pin down the
HVL~3 gain across the detector.  We note that these stellar
observations occurred in 1983 and 1985, so we checked the stability of
the UVS by comparing these observations to observations made in 1980
(but only in the normal stellar observation mode of pulse counting).
The differences were less than 5\%, consistent with the analysis of
post-Jupiter stability by \citet{Holberg82}.  However, stars are too
weak to use the HVL~2 gain, so we used special observations of the Sun
to relate the two gain states.  In these solar observations, which
occurred roughly a month before and after the Saturn occultations, the
UVS slit was slewed across the Sun in HVL~2 and then back across the
Sun in HVL~3.  The HVL~2 gain was determined by iteratively adjusting
it until the processed HVL~2 and HVL~3 spectra were as consistent as
possible.  These solar observations were also used to make small
corrections to the current limiting function of the UVS detectors,
necessary because the original function was determined for conditions
at Jupiter distances from the Sun.

A final and important element of the detector model is an empirical
correction for a problem with the even-numbered anodes of the
detector.  The problem manifests itself as a reduction in the counts
on even-numbered channels when the total signal on the detector
exceeds a certain threshold.  The primary effect is a factor of two
reduction in the number of channels that can be used to study the
atmosphere.  Although the exact cause of this anomaly is unknown, we were
able to use solar calibrations to develop a simple empirical
parameterization to correct for the effect and restore the even
channels to the appropriate counting levels.  This correction is
remarkably successful in reproducing the problem, and the first
application to Voyager UVS occultation data by \citet{Vervack97} and
\citet{Vervack04} allowed us to recover the full wavelength resolution
of the UVS during the Titan solar occultations.  Previous analyses
either neglected the even spectral channels or averaged the adjacent
odd channels, effectively halving the resolution.

With the above modeling approach we are not able to infer the proper
intensity in a channel that is saturated.  Fortunately, the only
occultation where saturation was an issue was the Voyager~1 solar
egress occultation, and the gain change eliminated this concern in all
but a small number of unused spectra (see above).

\paragraph*{Removal of scattered light.}
Modeling of scattered light is a difficult task; therefore, the
removal of instrumentally scattered light from the spectra was
accomplished through the application of a linear matrix based on
laboratory measurements (see \citet{Broadfoot81over} for a
description).  However, using in-flight data, we have made minor
changes to the matrix that were revealed as necessary in light of our
improved processing techniques.  These alterations are described by
\citet{Vervack97}.

It is important to note that the Monte Carlo model we employ is an
iterative forward-model method: we start with a guess at the photon
spectrum input to the UVS, generate the resulting counts spectrum via
our forward model of the detector system, compare the modeled counts
to the observed counts, and finally make changes to the input photon
spectra based on the differences.  This means the scattered light
matrix is applied in the forward direction and adds scattered light to
the modeled spectra for comparisons to what is actually observed.  The
foward scattering matrices are what were measured in the lab
pre-flight, and there are potentially issues regarding the accuracy of
the inverse versions of these matrices (i.e., the ones that would be
applied to remove scattered light from observed spectra) when applied
to bright solar observations.  The approach we utilize makes use of
the preferred matrices to provide the best accuracy in the resulting
processed spectra.

The use of this detector model in removing the instrumental effects
from UVS data significantly improves the accuracy of the spectral
processing.  In particular, corrections that have been neglected in
the past are now included.  The Monte Carlo approach allows for a
detailed calculation of the associated uncertainties in the resulting
spectra by including the inherent randomness of the charges generated
in the MCPs and propagating that randomness through the processing
procedure.  The resulting dataset of occultations is of higher
quality than previously available and is a primary factor in the
differences between our analysis and earlier analyses.

\subsection{Generation of Transmission Spectra}
\label{sec:transpec}

Generation of transmission spectra is conceptually simple: we just
need to divide the processed $I$ spectra by the $I_o$ spectra.
However, because the measured wavelength range and spectral intensity
of UVS spectra vary with the position of the source in the slit (see
section~\ref{sec:uvsdescrip}), we must account for the effects of the
limit cycle (attitude control) motions of the spacecraft.  It is
therefore critical that the $I$ and $I_o$ spectra be obtained at the
same relative position in the UVS slit for the $I/I_o$ spectra to
represent the true absorption of Saturn's atmosphere and not the
absorption convolved with observational circumstances.

Slit positions are characterized by two quantities: \dw\ and \dl.  The
primary quantity of importance is \dw, a measure of position along the
width (dispersion) direction, because the two effects of concern are
both connected to \dw.  The other position indicator \dl\ pertains to
the length direction of the slit, and the only effect along the length
of the slit is a variation of the detector sensitivity.  Because both
\dw\ and \dl\ are relative measures of the source position, it is
difficult to impossible to ascertain where along the length direction
the source is absolutely positioned.  Thus, although variations with
\dl\ are known to exist, it is generally not possible to correct for
them.  For the Saturn occultations, the motion in \dl\ is small
enough that the effects should be minimal, and we have neglected them.

Values of \dw\ for each $I$ and $I_o$ spectrum are calculated from
the roll, pitch, and yaw angle information in the engineering
telemetry.  Reference spectra are then binned using a \dw\ step size
of 0.001\degr.  This value is large enough to ensure acceptable
signal-to-noise in each bin but small enough to be only a fraction of
a spectral channel (each channel subtends 0.0286\degr) so that
spectral shifts from bin to bin are minimized.  Each bin is
normalized by the number of spectra in the bin to generate an average
reference $I_o$ spectrum at each bin-center \dw.  Finally, each $I$
spectrum is divided by the appropriate reference spectrum to generate
the final $I/I_o$ transmission spectra.

For the final step in generating transmission spectra, the
line-of-sight geometry for each spectrum is calculated.  Note that we
use radius and not altitude because Saturn is not a true ellipsoid but
rather has latitudinal variations superimposed on the bulk ellipsoidal
surface \citep{Lindal85}.  By providing results as a function of
radius, they are independent of any assumptions about the surface
shape.  To calculate the geometry, we adopt the method described by
\citet{Nicholson90}, in which all geometry calculations are performed
in a saturnocentric reference frame.  This approach allows the motion
of the source (Sun or star), Saturn, and the spacecraft to be taken
into account in a relatively simple fashion.  The necessary vectors
are determined using NAIF (Navigation and Ancillary Information
Facility) SPICE (Spacecraft Planet Instrument ``C-Matrix'' Events)
library routines obtained from NASA's Jet Propulsion Laboratory.  For
each spectrum, the vectors are determined using the ephemeris time of
the observation calculated from the spacecraft event time (SCET).  All
times and vectors are referenced to the J2000 inertial reference
frame that is used by the SPICE routines.  For details of the method
and the calculations, refer to \citet{Nicholson90} and
\citet{Vervack97}, respectively.

The uncertainties in the final light curves are generated by
propagating the uncertainties in the measured raw spectra through all
the spectral processing steps, reference spectra binning, and
transmission spectra generation using the standard methods for
propagation of uncertainties as outlined by \citet{Bevington92}.  They
are statistical in nature only (i.e., primarily the result of
channel-to-channel and spectrum-to-spectrum counting variations) and
do not account for systematic uncertainties that may be associated
with certain elements of the processing procedure (e.g., uncertainties
in the scattered light matrices are not known).  Fortunately, owing to
the use of tranmission spectra that are a ratio of two spectra, any
systematic uncertainties are likely minimized.

\section{Retrievals of Density Profiles}
\label{sec:method}

Two separate approaches have typically been adopted to infer
atmospheric density profiles in previous analyses of UVS occultation
data. In the first approach, a physically plausible atmosphere is
generated, the transmission spectra that would result from the model
are calculated, and adjustments to the atmospheric model are made
until the modeled transmission spectra provide the best possible match
to the measured spectra.  This approach was used in the original
Saturn occultation analyses of \citet{Smith83} and \citet{Festou82},
as well as in some modern-day occultation analyses of the giant
planets \citep[e.g.,][]{Greathouse10,Shemansky12}.  In the second
approach, mathematical inversion techniques are used to obtain direct
retrievals of the atmospheric densities from the transmission spectra
(e.g., the approach used in the Neptune occultation analysis of
\citet{Yelle93}).  The first ``forward-modeling'' approach has the
advantage that the resulting atmospheric solution is consistent with
physical laws, but the computational time can be long if many
iterations are required for convergence, and the results are nonunique
owing to unconstrained model input parameters.  The second
``mathematical-inversion'' approach is computationally more efficient
and provides better uncertainty estimation, but the method relies on
mathematical relationships that do not always approximate reality.

In this work, we employ the hybrid approach of \citet{Vervack04} 
\citep[see also][]{Koskinen13}.  As
in the first method, we use a forward-modeling technique to calculate
the transmission spectra, but rather than providing a fixed input
atmosphere, we rely on an iterative inversion technique to derive the
atmospheric densities directly from the data.  The main advantage of
this approach is that the transmission spectra can be accurately
modeled by including such effects as atmospheric attenuation,
wavelength blending (i.e., the measurement of many wavelengths in a
single detector channel), and, for solar occultations, the finite size
of the Sun in both the $I$ and $I_o$ spectrum {\em prior\/} to
dividing the two.  In contrast, the direct inversion approach requires
approximations of these effects (i.e., through parameterizations and
cross-section weightings) because the measured $I/I_o$ spectrum that
is the starting point of the inversions has the effects already
combined and inseparable.

The following discussion tracks that of \citet{Vervack04} but is
provided here in a sufficient amount of detail for a reader to
understand how the densities were retrieved.  For a more detailed 
description, see \citet{Vervack04}.  We retrieve densities through a 
two-step process.  In the first step, each transmission spectrum is
independently fit using the Marquardt-Levenberg minimization technique
described by \citet{Press92} to yield the line-of-sight column
densities.  The chi-squared spectral mismatch between the fitted and
observed transmission spectra is given by
\begin{equation}
  \Delta S = \sum_{c} { \left[ M(c) - D(c) \right]^2 \over \left[ \sigma_D(c)
    \right]^2} {\rm ,} \label{eq:chisq}
\end{equation}
\noindent
where $\Delta S$ is the mismatch function to be minimized, $M(c)$ and
$D(c)$ are the values of the model and observed transmission spectra
for channel $c$, $\sigma_D(c)$ is the uncertainty of the observed
transmission spectrum for channel $c$, and the summation is over the
$c$ channels of the spectra.  Once the spectral mismatch $\Delta S$ is
minimized, both the column densities and their associated covariance
matrix are returned by the Marquardt-Levenberg procedure.

In the second step, the line-of-sight column densities for each
species are inverted to yield the vertical number densities using a
constrained linear technique similar to that described by
\citet{Twomey77inv}.  We employ a second difference smoothness
constraint \citep[see][]{Twomey77inv} to provide a physical continuity
to the vertical profiles and assume a spherically symmetric atmosphere
with a 5-km radial grid spacing.  Linear quadrature is used to
describe the variation of the number densities within each layer, and
the profile falls off exponentially above the uppermost layer.  The
assumption of spherical symmetry is reasonably valid despite Saturn's
oblateness because the occultations generally probe along lines of
constant latitude where the geometry is approximately circular (as
opposed to elliptical).  The uncertainties associated with the final
vertical density profiles are calculated by propagating the column
density covariance matrix through the inversion
\citep[see][]{Menke89}.

Modeling the attenuated $I$ spectrum during the retrieval procedure is
a key step in the process.  We start with an input solar spectrum,
determine the line-of-sight optical depth $\tau$ at each wavelength,
and apply Eq.~\ref{eq:beer} to attenuate the spectrum.  The optical
depth $\tau$ is defined such that
\begin{equation}
  \tau(\lambda,r_{pca}) = \sum_i \int_{-\infty}^\infty \sigma_i(\lambda,s)
    \, n_i(s)\, ds {\rm ,} \label{eq:tauint}
\end{equation}
where $r_{pca}$ is the radial distance from the center of Saturn to
the point of closest approach along the line of sight, $n_i(s)$ is the
local number density of species $i$ along the line-of-sight direction
(denoted by $s$ with origin at the point $r_{pca}$), and
$\sigma_i(\lambda,s)$ is the total extinction cross section at
wavelength $\lambda$ along the line of sight.  In our case, the
$\sigma$ are simply photoabsorption cross sections.  The integral in
Eq.~\ref{eq:tauint} may be replaced with a summation over the layers
$j$ of our atmosphere to give
\begin{equation}
  \tau(\lambda,r_{pca}) = \sum_i \sum_{j_{low}}^{j_{max}}
    [\sigma(\lambda)]_{ij}\, N_{ij} {\rm ,} \label{eq:tausum}
\end{equation}
where $N_{ij}$ is the column density for species $i$ in layer $j$.  We
solve for the column densities directly, so we replace
Eq.~\ref{eq:tausum} with
\begin{equation}
  \tau(\lambda,r_{pca}) = \sum_i [\sigma(\lambda)]_i\, N_i {\rm ,}
    \label{eq:tausumtwo}
\end{equation}
where $N_i$ and $[\sigma(\lambda)]_i$ are the total line-of-sight column density (i.e., the sum of $N_{ij}$ over all $j$) and the
photoabsorption cross section for species $i$.  Although
$[\sigma(\lambda)]_i$ should in general be appropriately weighted to
account for line-of-sight temperature variations, cross section
measurements at temperatures relevant to Saturn's atmosphere over the
altitude range of the UVS occultations are few.  We therefore use
cross sections corresponding to the temperature at the point of
closest approach, where the largest relative amount of absorption occurs along the line of sight.  Our standard temperature profile (see
section~\ref{sec:temp}) is used to establish the cross sections to
use.

Variations in the temperature profile from occultation to occultation
are likely, but the variations in most cross sections with the likely
variations in temperature {\em at a given radius} are small enough
that we do not consider them here.  A more significant variation with
temperature is probable for \htwo\ absorption near the bottom of the
derived profiles where the temperature profile is changing rapidly
from the thermospheric value to the mesopause
value.  However, the majority of the \htwo\ profile is derived from
the region of nearly constant temperature in the thermosphere and
therefore is affected only to a small degree.  Even at the bottom of
the profile, runs assuming fixed temperatures of 100~K and 500~K
(bracketing the extremes of thermosphere and mesosphere) yield \htwo\
retrievals that differ only at the level of 10\% or less.

The SC\#21REFW reference spectrum of \citet{Hinteregger81} is used as
the input source spectrum in the solar occultation retrievals, taking
into account the heliocentric distance of Saturn.  Adjustments needed
to model the solar spectrum for the actual flyby dates are made using
appropriate factors as provided by K.~Fukui (personal communication,
1996).  Higher resolution spectra are available, but the
\citeauthor{Hinteregger81} spectrum was developed for the actual time
period covered by the Saturn occultations.  It is thus a more accurate
representation of the relative flux levels across the wavelength range
of the Voyager UVS, which has a more significant effect on the model
spectra owing to scattered light than the resolution does.  For the
stellar occultations, an absolutely calibrated average spectrum of the
star obtained from the unattenuated $I_o$ reference spectra region is
used for the source spectrum.  The sources of the cross sections used
in the retrievals are listed in Table~\ref{tab:photocs}; note that
only some of the species were included in the retrievals (further
discussed in section~\ref{sec:vtwosoling}).  Both the source spectra
and the cross sections were interpolated (appropriately weighted) onto
a 0.5~\AA\ grid prior to the calculations to ensure a small spacing in
the wavelengths of the input spectrum relative to the 9.26~\AA\ width
of the UVS spectral channels.  Further details of the construction of
both the source spectra and cross section datasets may be found in
\citet{Vervack97}.

\begin{table}[h]
  \caption{Sources for Photoabsorption Cross Section Data}
  \begin{tabular}{ll}
      \hline
      {\bf Species} & {\bf Source} \\
      \hline
      \htwo\ (continuum) & \citet{Samson94,Chan92} \\
                         & \citet{Ford73,Dalgarno69} \\
      \htwo\ (bands)     & \citet{Yelle93} \\
      H                  & \citet{Sadeghpour92,Samson66} \\
      \chfour\           & \citet{Au93,Samson89}; \\
                         & \citet{Mount78,Mount77} \\
      \ctwohtwo\         & \citet{Cooper95acet,Smith91}; \\
                         & \citet{Xia91,Suto84} \\
      \ctwohfour\        & \citet{Cooper95ethy,Zelikoff53} \\
      \ctwohsix\         & \citet{Au93,Mount78} \\
      \cfourhtwo$^{\rm a}$        & \citet{Okabe83} \\
      CH$_{\rm 3}$C$_{\rm 2}$H$^{\rm b}$        & \citet{Ho98} \\
      C$_{\rm 3}$H$_{\rm 6}$$^{\rm b}$        & \citet{Samson62,Orkin97} \\
      C$_{\rm 3}$H$_{\rm 8}$$^{\rm b}$        & \citet{Au93} \\
      \hline
  \end{tabular}

  $^{\rm a}$ \cfourhtwo\ was included in the Voyager~2 solar ingress and
             the Voyager~1 solar egress retrievals but profiles were not
             statistically significant and are not reported.

  $^{\rm b}$ These species were not included in the retrievals but were
             used in the test discussed in section~\ref{sec:vtwosoling} to
             demonstrate their exclusion.
  \label{tab:photocs}
\end{table}
\clearpage

After attenuation, each $I$ spectrum is degraded to UVS resolution.
We account for the UVS optical line shape, the slit function (an
improved version for the occultation port determined from a special
series of observations), and the limit-cycle-induced spectral shift.
For solar occultations, we also account for the finite size of the
Sun.  After applying these effects to each wavelength of the input
spectrum, the resulting spectrum is integrated over the wavelength
range of each spectral channel.  The $I_o$ spectrum is modeled in the
same fashion by degrading the unattenuated source spectrum, and the
ratio of the two spectra is taken to yield the transmission spectrum.

The results of our density retrievals for each occultation, along with
details particular to a given occultation, are presented below in
section~\ref{sec:results}.  In the next section, we discuss the
general method we employed for the generation of photochemical models
that simulate atmospheric conditions at each occultation location.

\section{Forward models}
\label{sec:forward}

The direct inversion technique has the benefit of providing the best
mathematical fit to the data.  However, the results can occasionally
be unphysical, such as when the derived methane mixing ratio appears
to increase with increasing altitude or when the ethane concentration
approaches the methane concentration in the lower part of the
occultation light curve for some of the noisier occultations (see
below).  The first situation is unphysical for a methane source from
deeper in the atmosphere, and our current understanding of methane
photochemistry on Saturn
\citep[e.g.,][]{Moses00,Moses05etal,Ollivier00,Dobrijevic03,Dobrijevic11}
precludes the second possibility.  Such problems can be avoided by using a
physics-based forward-modeling technique, but the forward models have
their own weaknesses, such as being poorly constrained and inherently
nonunique.  By using a combination of forward modeling and data
inversion, we can check for physical consistency and can exploit the
strengths of both methods to derive more realistic atmospheric
structure information for Saturn.  In this section, we add a forward
modeling component to the investigation in order to better determine
the implications of the UVS occultation results with regard to
chemistry and transport in Saturn's upper atmosphere during the
Voyager era.

The goal of the forward modeling is to reproduce the variation of
species densities with radius determined from the occultation light
curves and density retrievals and to make resulting inferences about
chemical and physical processes and properties in the homopause
region.  The \htwo\ density profile in Saturn's atmosphere depends on
the temperature structure, mean molecular mass, and gravity variation
with radius, none of which are well constrained below the regions
probed by the ultraviolet occultations. This is a significant yet
under-appreciated weakness of forward modeling that may be one cause
of the widely varying results on thermospheric temperatures from UVS
occultation analyses in the past \citep[cf.][]{Festou82,Smith83}, as
we discuss further in section~\ref{sec:stellardata} (see also Appendix
A).  Transforming the radius scale to a more meaningful pressure scale
for interpreting the underlying chemical and physical processes
occurring in the atmosphere is particularly difficult for the oblate
giant planets.  Saturn's unusual shape, rapid rotation, strong zonal
winds, uncertain radius at any reference pressure level as a function
of latitude, and uncertain mean molecular mass variation with altitude
complicate the derivation of the gravity and density structure.  We
often must extrapolate the densities from the upper troposphere, where
the planetary radius, zonal winds, and temperatures are better known
(and thus better constrain the density at a particular radius), to the
upper stratosphere and thermosphere, where the UVS occultations and
ground-based stellar occultations provide information.  In the
intervening region, temperatures, wind speeds, and other information
are uncertain, particularly during the Voyager encounter.  The Cassini
mission has provided a wealth of information on stratospheric
temperatures
\citep[e.g.,][]{Flasar05,Fletcher07,Fletcher08,Fletcher10,Fouchet08,Liming08,Li13,Guerlet09,Guerlet10,Guerlet11,Sinclair13,Sinclair14},
but the results show significant seasonal and interannual variability,
effectively preventing us from using Cassini information to
characterize stratospheric temperatures at the time of the Voyager
encounter.  Moreover, there is still a data gap between the lowest
pressure (highest altitude) that the Cassini Composite Infrared
Spectrometer (CIRS) data can probe during the limb observations
\citep[e.g.,][]{Guerlet09} and the pressures that the ultraviolet
occultations probe.

In any case, many plausible temperature profiles and other model
parameters can be considered that reproduce the \htwo\ density
structure derived from the inversion of the UVS occultation data ---
the forward-modeling solutions are simply nonunique.  It is also
possible that the \htwo\ is not in local thermodynamic equilibrium
\citep{Shemansky12}, resulting in poorly understood population states
that are difficult to model \citep[e.g.,][]{Hallett05rot}, affecting
the analysis of the occultation light curves in the \htwo\ Lyman and
Werner bands in particular.  We develop multiple plausible models for
several of the occultations to better illustrate the uncertainties in
the forward-modeling technique.

The variation of the hydrocarbon concentrations with radius depends
not only on the background atmospheric structure and \htwo\ density,
but also on chemistry and transport processes.  For the forward
modeling part of this investigation, we use the Caltech/JPL KINETICS
code \citep{Allen81} to develop one-dimensional models of
photochemistry and diffusion that define the variation of hydrocarbon
abundances with altitude on Saturn.  The adopted chemistry follows
that of Model~C of \citet{Moses05etal}.  We keep the models
consistent, to the highest extent possible, with Earth-based and
spacecraft observations concerning the stratospheric hydrocarbon
abundances below the region of sensitivity of the UVS occultations.
The key unknown input parameters in the photochemical models for our
purposes, aside from some uncertain reaction rate coefficients, are
the temperature profile, which controls the density variation with
altitude and can affect reaction rates, and the eddy diffusion
coefficient profile, which affects the variation of species
concentrations (and mean molecular mass) with altitude. Other
uncertainties such as the planetary shape and the zonal wind speed and
its variation with altitude contribute significant uncertainties to
the gravity field and the background atmospheric structure, which can
lead to systematic errors in the models.  The kinetic reaction rate
coefficients, photolysis cross sections, and other chemistry inputs
also have some effect on the results, but only for the complex
hydrocarbons.  Methane itself is controlled mainly by diffusion or
vertical transport in this portion of the atmosphere; chemistry has
only a minor influence on the \chfour\ profile.

We begin the forward modeling by specifying a radius at the 1-bar
level and a temperature and mean-molecular-mass profile as a function
of pressure from the 1-bar level on up to $\sim$3$\scinot-9.$ mbar.
The hydrostatic equilibrium equation is then solved to define a
background atmospheric grid with full pressure, temperature, radius,
gravity, and density structural information.  The graviational
acceleration is determined for a rapidly rotating, fluid planet as
described in \citet{Lindal85}.  We assume gravitational harmonic
coefficients $J_2$ = 0.016331, $J_4 = -0.000914$, and $J_6$ = 1.08$\,
\times \, 10^{-4}$ \citep{Nicholson88} and consider a
latitude-dependent rotation velocity that includes the zonal winds.
The latitude-dependent 1-bar radius and cloud-top zonal winds are
obtained from W. B. Hubbard (personal communication, 2004), and are
based on the \citet{Ingersoll84} wind fields and an assumption of a
strictly barotropic atmosphere, with wind speeds constant on a
cylindrical surface aligned with the rotation axis.
Table~\ref{tab:forward} describes the 1-bar radii and rotation rate
assumptions for our forward models.

\begin{table}[h]
\caption{Radius and Rotation-Rate Assumptions for Forward Models}
\begin{tabular}{lccc}
\hline
\           & Planetocentric & Assumed Radius & Local Rotation\\  
Occultation & Latitude       & at 1 bar (km)  & Period (hr) \\
\hline 
Voyager 2 solar ingress    & $29\deg$    & 58556 & 10.60 \\
Voyager 2 stellar egress   & $3.8\deg$   & 60232 & 10.18 \\
Voyager 1 stellar egress   & $-4.8\deg$  & 60213 & 10.22 \\
Voyager 2 stellar ingress  & $-21.5\deg$ & 59260 & 10.41 \\
Voyager 1 solar egress     & $-27\deg$   & 58771 & 10.56 \\
\hline
\end{tabular}
\noindent{Note: Our solution to the hydrostatic equilibrium equation 
requires knowledge of the local angular velocity (i.e., the angular 
velocity corresponding to the interior rotation period plus the angular
velocity due to the zonal wind).  As long as the winds were determined
in a consistent manner relative to the assumed interior rotation period,
our solution is independent of the exact interior rotation period --- it
is the local atmospheric rotation period that is used in the equations.}
\label{tab:forward}
\end{table}
\clearpage

For lack of better information, we assume that the zonal winds derived
at the cloud tops are constant with height.  Note that this assumption
differs from our data inversion described in the previous sections, in
which we assume that zonal winds have dropped to zero at these high
altitudes, so some differences exist between the
density-temperature-radius profiles derived from the foward modeling
and inverse retrievals.  \citet{Hubbard97} provide evidence for the
zonal winds in the equatorial region being strong up to at least the
microbar region.  However, recent Cassini and ground-based data show
some evidence that Saturn's zonal thermal wind structure contains
vertical shear, especially in the near-equatorial region where there
appears to be a kind of $\sim$15-year oscillation similar to the
Earth's semi-annual oscillation
\citep{Fouchet08,Orton08,Liming08,Li11,Guerlet11}.

\begin{figure}[p]
    \includegraphics[scale=0.59,angle=-90]{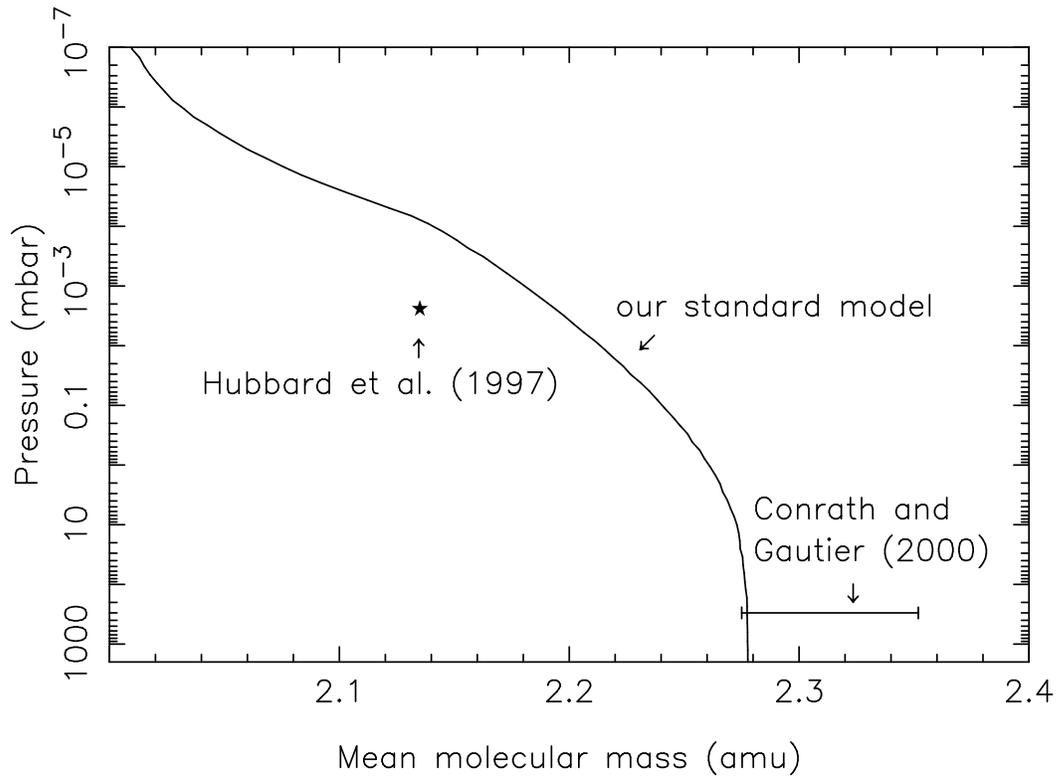}
    \caption{The mean molecular mass variation with pressure, as
    derived from our standard photochemistry/diffusion model with an
    assumed He mole fraction of 10\% at the 5-bar lower boundary
    (solid line), is compared with the Voyager IRIS derivation (range
    with error bars) of Conrath and Gautier (2000) and the 2.135 amu
    assumption (star) of Hubbard et al. (1997).  Note that although we
    expect the mean molecular mass to drop off with altitude, the
    exact profile will depend on uncertain vertical transport
    parameters.}
    \label{fig:mass}
\end{figure}

The preliminary background hydrostatic atmosphere is then combined
with an eddy diffusion coefficient profile derived from previous
photochemical modeling \citep{Moses05etal} and is used to develop a
new preliminary photochemical model for the equatorial region.  The
resulting abundance profiles are used to derive a mean molecular mass
profile (see Fig.~\ref{fig:mass}) and a refractivity versus radius
profile for comparison with the \citet{Hubbard97} stellar occultation
data (reported for the equatorial region).  In the case of a poor fit
to the Hubbard et al.~refractivity values, the temperature profile is
tweaked and the entire process iterated, until the
\citeauthor{Hubbard97} refractivity profile in their 60830-61040 km
region is reproduced.  Note that we elect to compare with the
\citeauthor{Hubbard97} refractivity profile rather than their derived
temperature profile because of differences in our assumptions about
the atmospheric composition and mean molecular mass.  At the lower
boundary of our model (5 bar), we assume a helium mole fraction of
10\%.  This value is at the lower end of the 12$(\, \pm \, 2)$\% range
derived by \citet{Conrath00} from an analysis of Voyager IRIS spectra
but is higher than the combined Voyager IRIS-RSS derivation of 3.4$(\,
\pm \, 2.4)$\% \citep{Conrath84} and higher than the preliminary
combined Cassini RSS-CIRS derivation of $\sim$7\%
\citep{Flasar08,Fouchet09} or the \citet{Hubbard97} assumption of 6\%.
The tropospheric helium abundance on Saturn remains uncertain.
Throughout the modeling, we also assume a \chfour\ mole fraction of
4.5$\scinot-3.$ \citep{Flasar05} at the lower boundary (5~bar) of our
models \citep[cf.][]{Fletcher09}.

\clearpage

\subsection{Temperature structure adopted in the forward models}
\label{sec:temp}

The thermal structure of Saturn is well enough understood to define
three distinct regions based on the temperature profile: (1) the
troposphere, below the temperature minimum at $\sim$100 mbar, in which
radiative-convective exchange controls the temperatures, (2) the
middle atmosphere, between $\sim$100 and a few $\times$ 10$^{-5}$
mbar, in which radiative processes control temperatures and in which
the atmosphere is stable against convection, and (3) the thermosphere,
above a few $\times$ 10$^{-5}$ mbar (but depends on latitude and/or
time), in which the temperature begins to increase dramatically with
altitude up to a nearly isothermal exospheric region.  The heating
mechanisms for the thermosphere are currently not well understood
\citep[e.g.,][]{Yelle04,MullerWodarg06,MullerWodarg12,Smith07,Smith08}.
Throughout this paper, we use the term ``stratosphere'' to refer to
the entire middle atmosphere, and the term ``mesopause'' to refer to
an ill-defined region marking the base of the thermosphere.

Saturn's stratosphere above $\sim$0.01 mbar and the thermosphere are
difficult regions to access from remote-sensing methods.  The three
techniques of (1) Earth-based observations of stars being occulted by
Saturn \citep[e.g.,][]{Hubbard97,Cooray98,French99,Harrington10}, (2)
spacecraft ultraviolet observations of stars or the Sun being occulted
by Saturn \citep[e.g.,][this
  work]{Sandel82eddy,Festou82,Smith83,Feng91,Feng05,Shemansky12,Koskinen13},
and (3) observations of the planet's airglow and auroral emission
\citep[e.g.,][]{Sandel82sat,McGrath92,Emerich93,Geballe93,BenJaffel95,
  Gerard95,Gerard04,Gerard09,Parkinson98,Miller00,Hallett05uvis,Melin07,Melin11,Gustin09,Stallard12,Odonoghue14}
have provided our only constraints on the atmospheric
density/temperature structure in Saturn's thermosphere and mesopause
region.  However, deriving temperatures and densities from these
methods requires several assumptions about atmospheric properties that
are not well known.  For the forward modeling part of this
investigation, we consider the upper atmospheric temperature to be a
free parameter, and we adjust the temperature profile until we find a
good fit to the \htwo\ density profile derived from our inversion of
the UVS occultation data.

Although we started with temperature profiles consistent with the
\citet{Lindal85} Voyager radio occultation results and/or the Voyager
IRIS analysis of \citet{Conrath00}, we found that we needed to
increase temperatures in the middle atmosphere and/or troposphere in
order to match either the \citet{Hubbard97} refractivity data or the
UVS \htwo\ density inversions (see below), given our assumptions about
the 1-bar radius and the mean molecular mass.  Our warmer profiles are
consistent with the temperatures derived for late spring and early
summer latitudes from current-day mid-infrared observations
\citep[e.g.,][]{Greathouse05strato,Flasar05,Fletcher07,Guerlet09} but
are warmer than has been typically assumed or derived from
observational analyses from the Voyager era or Infrared Space
Observatory (ISO) era
\citep[e.g.,][]{Lindal85,Conrath00,Moses00,Lellouch01}.  Although our
result is dependent on initial assumptions, especially with regard to
the 1-bar radius and the helium abundance, we have developed numerous
models and covered enough parameter space to form the general
conclusion that the colder stratospheric temperatures derived from the
analysis of the global-average ISO data \citep{Moses00,Lellouch01} are
too cold to be accommodated by our UVS data-model comparisons --- the
stratosphere at 29\deg\ N latitude at the time of the Voyager 2 solar
ingress occultation was warmer and more expanded in an average sense
than is indicated by the ISO analysis (see the journal supplemental
material and section~\ref{sec:stellardata} for the forward-model
profiles).  It is difficult to quantify the temperature uncertainty
from the forward modeling, but one should not place too much reliance
on the actual assumed profile.  The ``average'' temperature below the
occultation altitudes needs to be relatively warm to account for the
\htwo\ abundance at large radial distances, but neither the vertical
profiles nor the location and temperature of the mesopause are
uniquely constrained by the comparisons of our forward models with the
UVS data.

\section{Results and Discussion: Comparison of Retrievals and Models}
\label{sec:results}

In this section we present the results of both our retrievals (data
inversions) and our photochemical models (forward models).  We begin
with the occultation of highest quality, the Voyager 2 solar ingress,
and proceed through the occultations in order of decreasing quality.
For each occultation, we present density profiles from both the
retrievals and models, as well as comparisons of the measured light
curves to light curves simulated using both the retrieved and model
densities.  For each occultation, we discuss the implications and
limitations of the comparisons, as well as the details that are
particular to each of the occultations.

\subsection{Voyager 2 Solar Ingress}
\label{sec:vtwosoling}

\subsubsection{Data Retrievals}

Owing to the high signal-to-noise ratio that results from using the
Sun as a source and to problems with the Voyager~1 solar occultations
(see below), the Voyager 2 solar ingress occultation represents the
best dataset available from the Voyager UVS occultations at Saturn,
and the profiles are of excellent quality.

\begin{figure}[p]
    \includegraphics[width=6.5in]{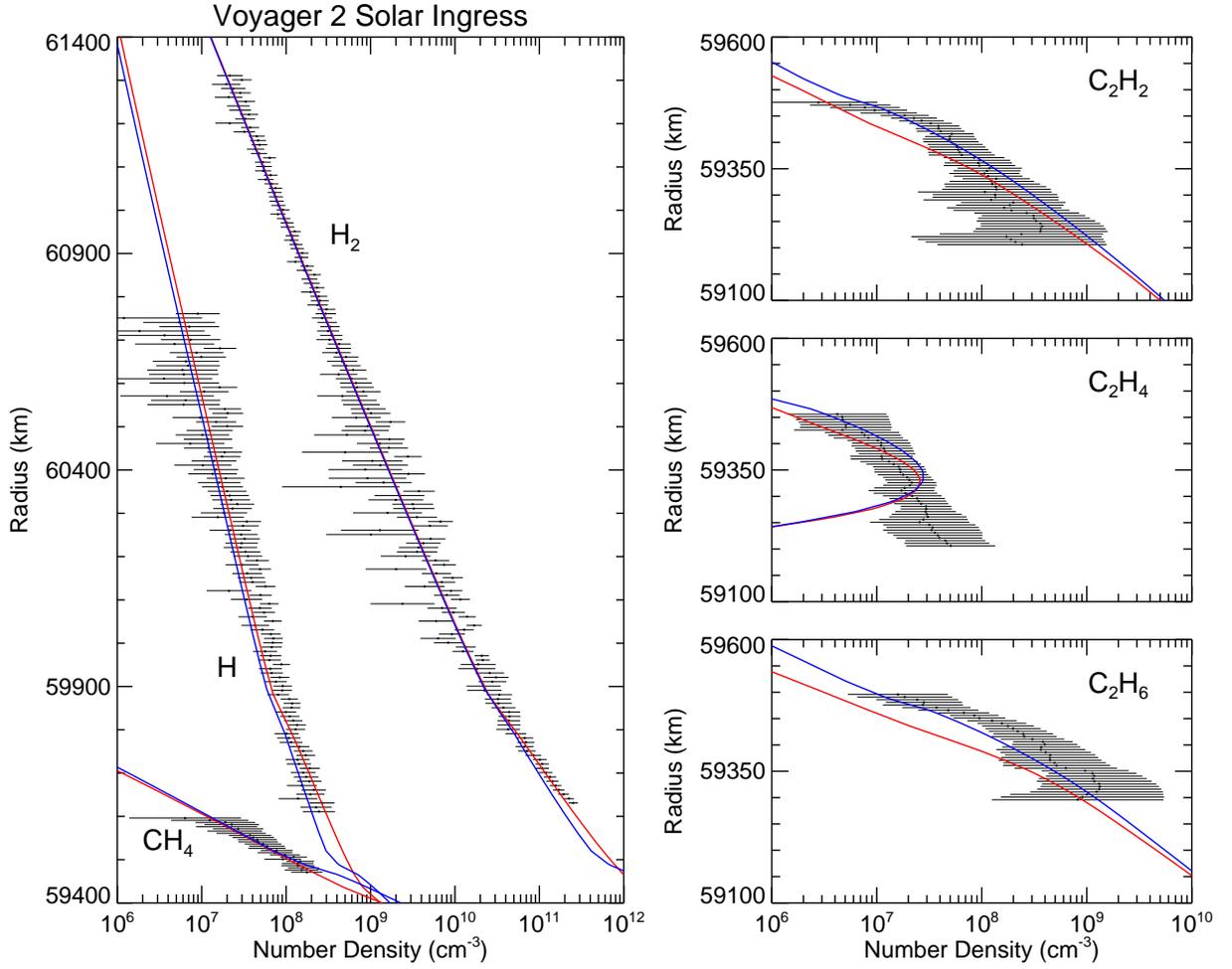}
    \caption{The density profiles from our standard model (red curve)
    and ``best-fit hydrocarbon model'' (blue curve) are compared with the
    densities retrieved from the data inversions of the Voyager~2
    solar ingress occultation at 29\deg\ latitude (black symbols, with
    associated one-sigma uncertainties).  The left figure is for
    \htwo, H, and \chfour, while the right panels are for \ctwohtwo\
    (top), \ctwohfour\ (middle), and \ctwohsix\ (bottom).  Only every
    other point is shown for the \htwo\ and H profiles for clarity.}
    \label{fig:vtwosolingret}
\end{figure}

Figure~\ref{fig:vtwosolingret} shows the density profiles retrieved
from the Voyager~2 solar ingress occultation.  For clarity, only every
other point is shown for \htwo\ and H.  Note that the hydrocarbon
profiles were retrieved using a tighter smoothing constraint than was
used for the \htwo\ and H profiles, which explains their smoother
nature relative to the profiles for \htwo\ and H.  This smoothing was
necessary because of the higher noise levels in the transmission
spectra at the wavelengths of hydrocarbon absorption.  In this and all
similar figures for the other occultations, the error bars on the
density profiles represent the one-sigma level.

\clearpage

The jittery nature near the middle of the \htwo\ profile represents
the altitude region where the density information transitions from the
\htwo\ continuum to the \htwo\ bands.  Because the continuum region is
almost completely absorbed whereas the band region is only beginning to
show absorption, the retrieved profile exhibits more variation than at
altitudes above and below.  This ``noisiness'' is consistent with the
fact that the retrievals at these altitudes are derived from data that
are at the extremes of transmission (i.e., either small residuals near
complete absorption or small absorptions near complete transmission).
Note that the \htwo\ profile shows just a hint of curvature at the
lowest altitudes, indicating that the temperature is changing to
smaller values.  This result is consistent with the earlier analysis
of \citet{Smith83} and, as we will see in section~\ref{sec:further},
with the model of \citet{Hubbard97}.

\begin{figure}[p]
    \includegraphics[width=6.5in]{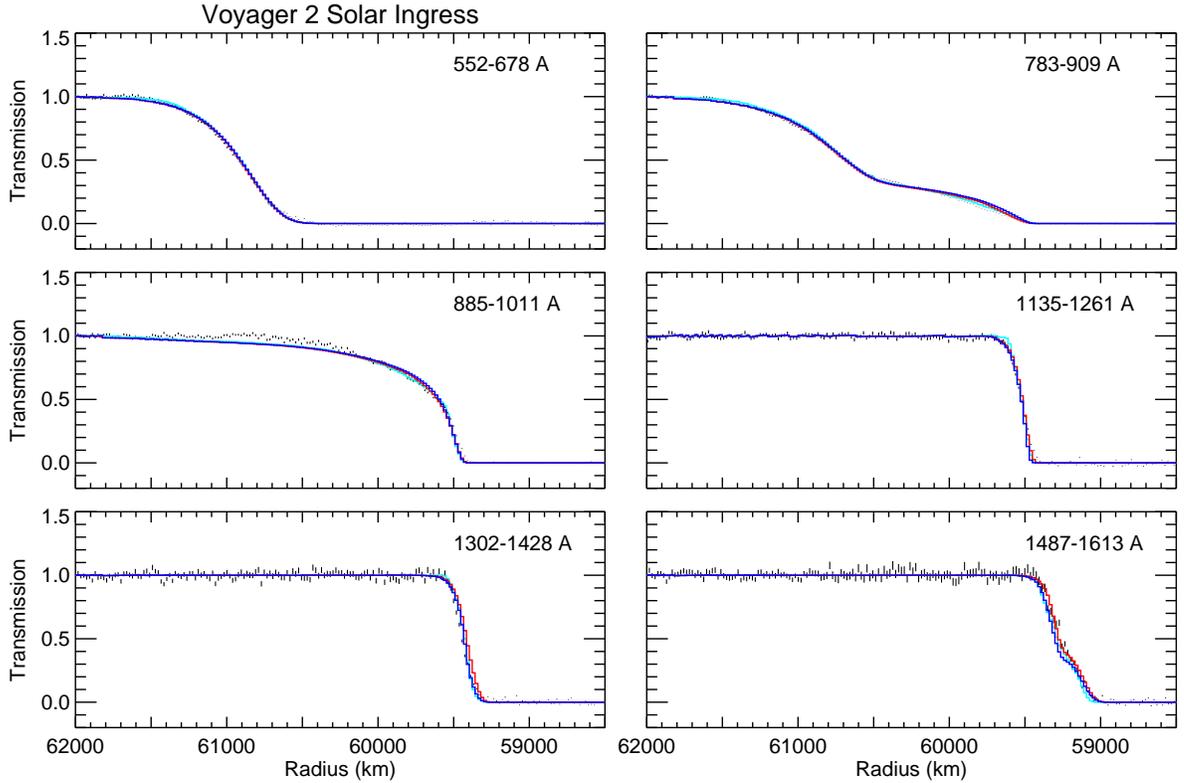}
    \caption{Light curves from our reanalysis of the Voyager~2 UVS
    solar ingress occultation (black bars representing the one-sigma
    uncertainty levels at each point) compared with synthetic light
    curves generated from our ``standard'' photochemical model (red)
    and ``best-fit hydrocarbon'' model (blue).  A synthetic light
    curve is also shown for the retrieved densities (light blue).
    These curves are averages of several channels and span the
    wavelength ranges indicated in each panel (including
    limit-cycle-induced shifts).  Only every third point is shown for
    clarity.  The top two panels represent absorption due primarily to
    \htwo\ and H.  The middle left panel is dominated by \htwo\
    absorption at higher radii and \chfour\ at lower radii, while the
    middle right panel is dominated by \chfour\ absorption at all
    altitudes.  The bottom two panels show aborption largely caused by
    complex hydrocarbons.}
    \label{fig:vtwosolinglc}
\end{figure}

Figure~\ref{fig:vtwosolinglc} shows a series of light curves to
demonstrate how well the retrieved densities reflect the observed
transmission.  In each panel, a data light curve representing the
uncertainty-weighted average of the light curves probing the specified
wavelength range (including shifts induced by $\Delta W$ motion) is
shown, and the corresponding synthetic light curve generated using the
retrieved density profiles is overplotted (light blue curve).  The
one-sigma uncertainties shown for the data curves represent the simple
average uncertainty over the same range rather than the weighted
uncertainty.  The average uncertainty is used to give the reader a
better feel for the uncertainties in the individual light curves that
were averaged to generate the shown data curve.  The weighted
uncertainties would, of course, be smaller.  The relatively small size
of the uncertainties indicates the high quality of the Voyager~2 solar
occultation.

\clearpage

Owing to channel-to-channel noise in the data, synthetic light curves
resulting from the retrieved density profiles can be poor fits on an
individual channel basis.  Because the retrieved profiles result from
spectral fits across many wavelength channels, it is better to compare
against an averaged light curve such as those we show here.  Note that
the level of the observed light curves exceeds one for some data
points because these curves represent the ratio of two measured
spectra.  The average level along the light curves is generally equal
to or less than one, but any given point may be greater than one
because of uncertainties in the measured spectra.

The wavelength ranges were chosen to span a variety of absorption
regimes.  The upper left panel is dominated by \htwo\ continuum
absorption.  The upper right panel is a combination of \htwo\ and H
continuum absorption with some \htwo\ band absorption.  The middle
left panel is dominated by \htwo\ band absortption for the most part
before absorption by \chfour\ becomes important at the end.  The
middle right panel is primarily represents absorption by \chfour.  The
lower left and lower right panels are dominated by absorption from
\ctwohtwo, \ctwohfour, and \ctwohsix.  As can be seen, the structure
in the data light curve in each panel is fit well by the synthetic
light curve (light blue curve in Fig.~\ref{fig:vtwosolinglc}).  The
worst fit is obtained for the \htwo\ band region shown in the middle
left panel.  This mismatch is possibly a reflection of small changes
being needed in the temperature profile used in the retrievals or
perhaps small changes in the modeled \htwo\ band absorption cross
sections.  Nevertheless, the overall fit is still excellent and the
deviations are relatively minor.  Tests done to examine the
sensitivity of the retrievals to changes in temperature and therefore
the \htwo\ band absorption show that the deviations between the data
and synthetic light curves represent less than a 5\% change in
density.

\subsubsection{Photochemical Modeling}

Because the Voyager 2 solar ingress occultation had the highest
signal-to-noise of all the Voyager UVS occultations, we have devoted a
major part of our effort to modeling this particular occultation.  We
present two different forward models for this occultation that both
provide a good fit to the retrieved \htwo\ and \chfour\ density
profiles and the occultation lightcurves.
%The temperature profiles for both models are shown in Fig.~\ref{fig:temp}.
Both models assume the ``Model C'' chemistry of \citet{Moses05etal}.
Our standard model has an eddy diffusion coefficient profile derived
from the following expression:
$$
%\halign to \hsize{$#$\hfill & $#$\hfill & $#$\hfill\quad \cr
\halign to \hsize{$#$\hfill & $#$\hfill & $#$\hfill\quad & $#$\hfill\quad \cr
K_{zz} & \ = \ & 1.6\scinot7.\, (2.0\scinot-4./p)^{0.85} & \ \ {\rm For}\  p < 2.0\scinot-4.\ \hbox{mbar} \cr
%\ &   &  \ \ \ {\rm For}\  p < 2.0\scinot-4.\ \hbox{mbar} \cr
\ & \ = \ & 1.6\scinot7.\, (2.0\scinot-4./p)^{0.79} & \ \ {\rm For}\ 2.0\scinot-4. \le p < 0.5\ \hbox{mbar} \cr
%\ &   &  \ \ \ {\rm For}\ 2.0\scinot-4. \le p < 0.5\ \hbox{mbar} \cr
\ & \ = \ & 3.3094\scinot4.\, (0.5/p)^{0.78} & \ \ {\rm For}\ 0.5 \le p < 100\ \hbox{mbar} \cr
%\ &   &  \ \ \ {\rm For}\ 0.5 \le p < 100\ \hbox{mbar} \cr
\ & \ = \ & 5.0\scinot2. & \ \ {\rm For}\ 100 \le p < 640\ \hbox{mbar} \cr
%\ &   &  \ \ \ {\rm For}\ 100 \le p < 640\ \hbox{mbar} \cr
\ & \ = \ & 2.5\scinot4. & \ \ {\rm For}\ p > 640\ \hbox{mbar} \cr}
%\ &   &  \ \ \ {\rm For}\ p > 640\ \hbox{mbar} \cr}
$$
for eddy diffusion coefficient $K_{zz}$ in cm$^2$ s$^{-1}$ and
pressure $p$ in mbar.  As is shown in Fig.~\ref{fig:vtwosolingret},
this model (red curves) provides a good fit to the retrieved \htwo\
and \chfour\ density profiles but not to the profiles for the
C$_2$H$_x$ hydrocarbons.  The other model (blue curves), dubbed the
``best-fit hydrocarbon'' model, provides a statistically better fit to
the hydrocarbon data but not to the \htwo\ density data.  In this
``best-fit hydrocarbon'' model,
$$
%\halign to \hsize{$#$\hfill & $#$\hfill & $#$\hfill\quad \cr
\halign to \hsize{$#$\hfill & $#$\hfill & $#$\hfill\quad & $#$\hfill\quad \cr
K_{zz} & \ = \ & 1.56\scinot7.\, (2.0\scinot-4./p)^{0.87} & \ \ {\rm For}\  p < 2.0\scinot-4.\ \hbox{mbar} \cr
%\ &   &  \ \ \ {\rm For}\  p < 2.0\scinot-4.\ \hbox{mbar} \cr
\ & \ = \ & 1.56\scinot7.\, (2.0\scinot-4./p)^{0.79} & \ \ {\rm For}\ 2.0\scinot-4. \le p < 0.5\ \hbox{mbar} \cr
%\ &   &  \ \ \ {\rm For}\ 2.0\scinot-4. \le p < 0.5\ \hbox{mbar} \cr
\ & \ = \ & 3.2267\scinot4.\, (0.5/p)^{0.78} & \ \ {\rm For}\ 0.5 \le p < 100\ \hbox{mbar} \cr
%\ &   &  \ \ \ {\rm For}\ 0.5 \le p < 100\ \hbox{mbar} \cr
\ & \ = \ & 5.0\scinot2. & \ \ {\rm For}\ 100 \le p < 640\ \hbox{mbar} \cr
%\ &   &  \ \ \ {\rm For}\ 100 \le p < 640\ \hbox{mbar} \cr
\ & \ = \ & 2.5\scinot4.  & \ \ {\rm For}\ p > 640\ \hbox{mbar} \quad . \cr}
%\ &   &  \ \ \ {\rm For}\ p > 640\ \hbox{mbar} \quad . \cr}
$$
Three other changes were made to the ``best-fit hydrocarbon model'': (1)
the low-pressure limiting rate constant for the reaction CH$_3$ +
CH$_3$ + M was increased, using the expression recommended by \citet{Smith03},
(2) the adopted molecular diffusion coefficients for the
stable C$_2$H$_x$ hydrocarbons \citep[see][]{Moses00} were reduced
by 30\%, and (3) the constraint for reproducing the \citet{Hubbard97}
refractivity profile was relaxed when constructing the model
temperature profile.

Figure~\ref{fig:vtwosolingret} shows how the density profiles from the
two photochemical models compare with the retrievals, whereas
Fig.~\ref{fig:vtwosolinglc} compares synthetic light curves created
from the photochemical model results with the observed light curves.
The standard model does an excellent job of reproducing the \htwo\ and
\chfour\ density profiles, as well as reproducing the observed light
curves in the regions where the absorption is dominated by \htwo\ and
\chfour\ (the top four panels of Fig.~\ref{fig:vtwosolinglc}).  The
standard model accurately predicts the \ctwohtwo\ and
\ctwohfour\ abundances near the midpoint altitude of the retrieved
profiles but underpredicts the \ctwohsix\ density at most altitudes,
and does not accurately predict the slopes of the profiles.  Although
the data retrieval process may be misassigning some of the
contributions (e.g., some of the absorption attributed to
\ctwohfour\ may actually be due to \ctwohtwo, and/or some of the
absorption attributed to \ctwohsix\ may actually be due to \chfour),
the light curve comparisons in Fig.~\ref{fig:vtwosolinglc} suggest
that the standard model has slightly insufficient hydrocarbon
absorption, at least at the upper altitudes of the retrievals.  The
``best-fit hydrocarbon'' model does a better job reproducing both the
retrieved densities and the light curves in the regions dominated by
the heavier hydrocarbons (lower two panels) although the model still
slightly underpredicts the C$_2$H$_x$ hydrocarbon absorption at high
altitudes and underpredicts the hydrocarbon absorption at the lowest
altitudes extracted from the retrievals.

These model-data comparisons will aid photochemical modelers in
constraining model inputs or otherwise interpreting the observed
behavior.  For example, the rate coefficients for the relevant
hydrocarbon reactions are not well known at low pressures (and
temperatures), and it is not surprising that the slopes of the model
profiles do not exactly reproduce the data.  Both the retrieved slopes
and the overall retrieved \ctwohtwo/\ctwohfour/\ctwohsix\ ratio will
help modelers distinguish the dominant chemical pathways occuring in
the homopause region of Saturn and the other giant planets
\citep[e.g.,][]{Moses05etal,Nagy09,Fouchet09}.

\begin{figure}[p]
    \includegraphics[scale=0.75]{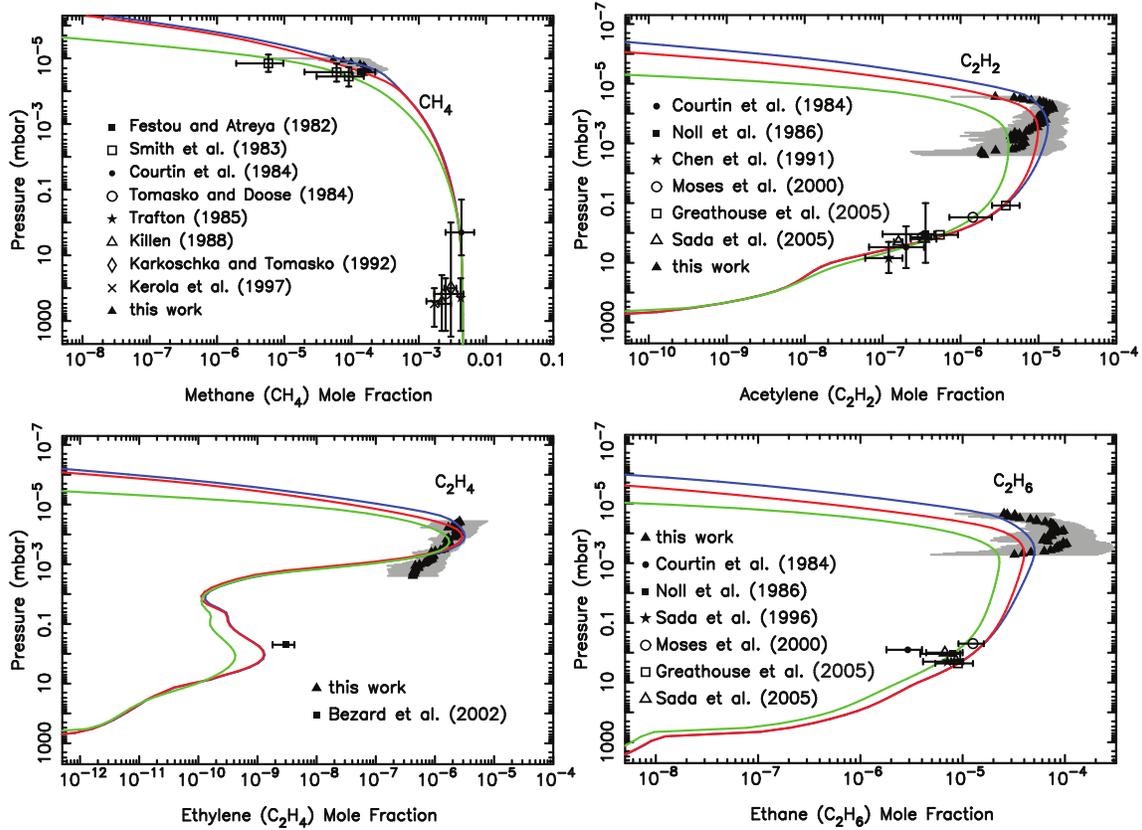}
    \caption{The mixing ratios of \chfour\ (top left), \ctwohtwo\ (top
      right), \ctwohfour\ (bottom left), and \ctwohsix\ (bottom right)
      in our standard model (red) and ``best-fit hydrocarbon'' model
      (blue), as compared with the retrievals from the Voyager~2 solar
      ingress occultation (solid triangles) and various other
      pre-Cassini observations (as labeled).  The pressure scale for
      the UVS retrievals plotted here is taken from the
      radius-vs-pressure grid from the ``best-fit hydrocarbon'' model
      and is very model dependent.  The green curve is Model C from
      \protect\citet{Moses05etal}.
      \protect\nocite{Courtin84,Tomasko84,Trafton85,Noll86,Killen88,Chen91,Karkoschka92,Sada96,Kerola97,Bezard02,Sada05}}
    \label{fig:vtwosolingmix}
\end{figure}

Additional insight into the implications for chemisty/transport in
Saturn's homopause region can be gained from looking at the mixing
ratio profiles as a function of pressure (see
Fig.~\ref{fig:vtwosolingmix}).  Note that the conversion from a
density-radius profile, which is the primary product from the UVS data
inversion, to a mixing ratio-pressure scale is problematic.  The main
problem is that the ultraviolet wavelengths that are sensitive
primarily to \htwo\ extinction become completely absorbed by the time
the longer-wavelength \chfour\ absorption becomes significant, so that
no overlap exists between the radius levels for the \htwo\ and
\chfour\ density retrievals.  That complicates the derivation of both
pressures and mixing ratios.  Converting from radius to pressure is
thus very model dependent, with the temperature profile being the
biggest uncertainty in the models.  For Fig.~\ref{fig:vtwosolingmix}, we
have converted radius to pressure using our model grid from our
``best-fit hydrocarbon'' model.

\clearpage

Although a detailed and thorough discussion of the implications of our
model-data comparisons with regard to chemistry and transport in
Saturn's homopause will be deferred to future work, we can make a few
general statements regarding the comparisons.
Figure~\ref{fig:vtwosolingmix} illustrates that methane and the C$_2$H$_x$
hydrocarbons are being carried to very low pressures (high altitudes)
in Saturn's atmosphere at 29\deg\ latitude at the time of the
Voyager~2 solar ingress occultation, implying strong vertical mixing
or upward winds in Saturn's upper atmosphere.  This result is
qualitatively in accord with previous analyses from various Voyager
UVS data
\citep[e.g.,][]{Sandel82eddy,Festou82,Atreya82,Smith83,Atreya84,BenJaffel95,Parkinson98}.
A direct quote of the value of $K_{zz}$ at the homopause in our model
is not very helpful here because our assumed slope of $K_{zz}$ at high
altitudes is so large (e.g., in the extreme case of our ``best-fit
hydrocarbon'' model, the official homopause level would be at
$\sim$7$\scinot-7.$ mbar, with a $K_{zz}$ of $\sim$2$\scinot9.$ cm$^2$
s$^{-1}$).  A better comparative measure is the value of $K_{zz}$ at
the half-light point for the occultation at the wavelengths for which
methane absorption dominates the light curve.  For the Voyager~2 solar
ingress occultation at 29\deg\ latitude, that radius is $\sim$59463~km
(907~km above the 1-bar radius).  The corresponding pressure in our
``best-fit hydrocarbon'' model is $\sim$2.4$\scinot-5.$ mbar, at which
point $K_{zz}$ $\approx$ 1.0$\scinot8.$ cm$^2$ s$^{-1}$.  In our
standard model, the corresponding pressure is $\sim$3.1$\scinot-5.$,
at which point $K_{zz}$ $\approx$ 7.9$\scinot7.$ cm$^2$ s$^{-1}$.  As
can be seen from Fig.~\ref{fig:vtwosolingmix}, our analysis suggests that
atmospheric vertical mixing at 29\deg\ latitude at the time of the
Voyager~2 UVS solar ingress occultation was as strong or even stronger
than has been inferred from the previous UVS analyses for the
Voyager~2 stellar egress occultation of $\delta$ Sco
\citep{Festou82,Smith83} at 3.8\deg\ latitude.

The implications with regard to chemistry in the homopause region are
harder to pin down, and more modeling will be required to fully
investigate this issue.  The standard model appears to have
insufficient hydrocarbon absorption at the highest altitudes probed by
the occultations.  The fit cannot be improved by simply altering some
of the uncertain chemical reaction rate coefficients from the
\citet{Moses05etal} model, as the molecules are in the
diffusion-dominated regime at these pressures.  Changes to chemistry
inputs have little effect at these low pressures.  The molecular
diffusion coefficient for methane in \htwo\ and He has been measured
\citep[e.g., see the review of][]{Marrero72}. \citet{Moses00} used
strategies outlined in \citet{Reid87} to scale the expression for
methane to derive molecular diffusion coefficients for the other
hydrocarbons, and we follow those expressions.  If we arbitrarily
assume a $\sim$30\% error for these expressions and reduce the
C$_2$H$_x$ molecular diffusion coefficients by 30\%, as we did in our
``best-fit hydrocarbon'' model, we can increase the hydrocarbon
densities at high altitudes.  Alternatively, we could have modified
the slope of the temperature profile going into the thermosphere to
obtain a better fit, or assumed that the profiles are controlled by an
upward vertical wind.

Again, non-uniqueness problems affect the forward-modeling analysis, as
it is possible to tweak several model free parameters to get the
desired answer.  However, we can safely say that the hydrocarbon
model-data mismatch at the highest altitudes is likely due to
transport effects or the temperature structure being inaccurately
reproduced in the model rather than being due to chemical processes.
These slight model-data mismatches do not detract the general firm
conclusion of vigorous mixing in Saturn's upper atmosphere at this
particular location and time.

Diffusion time scales in the models are greater than a saturnian day
at all pressures greater than a few $\times 10^{-6}$ mbar.  Chemical
loss time scales for the observed hydrocarbons are also greater than a
saturnian day at the observed altitudes, but the chemical loss time
scales are less than a saturnian season.  Therefore, we do not expect
any diurnal variations in insolation or diffusion to affect the
modeled abundances.  However, sufficiently strong diurnally variable
winds might affect the species profiles (see
section~\ref{sec:remarks}), and seasonal variations in the species
profiles and vertical winds are expected based on seasonal variations
in solar insolation \citep[see][]{Moses05great,Friedson12}.

In the lower portion of the density retrievals, the models again
depart from the abundances derived from the inversions.  The most
obvious departure is for \ctwohfour, for which the models apparently
greatly underpredict the \ctwohfour\ abundance at the lowest points of
the occultations.  This model-data mismatch occurs below the main
hydrocarbon production region in which H Lyman $\alpha$ photolysis of
\chfour\ dominates the chemistry; the problem may be due to incorrect
photolysis quantum yields for methane or the other hydrocarbons at
wavelengths longer than H Lyman $\alpha$, may be due to an incorrect
description of reactions that convert \ctwohtwo\ to \ctwohfour\ at
these altitudes, or may be due to some other problem with the adopted
chemistry.  A similar model-data mismatch occurs for the other Voyager
UVS occultations, as described below.  Furthermore, \citet{Nagy09}
find that the same type of model-data mismatch occurs when models
developed with the same basic chemistry inputs are compared to
hydrocarbon profiles derived from the preliminary Cassini UVIS
occultation analysis (\citet{Shemansky05}; see also
\citet{Shemansky12}), lending credence to the suggestion that there
are problems with the adopted \ctwohfour\ chemistry in the models
below the H Lyman $\alpha$ production peak.  As is consistent with the
discussion in \citet{Nagy09}, the chemical reaction list for Model A
of \citet{Moses05etal} does a better job of reproducing the slopes of
the C$_2$H$_x$ hydrocarbons as compared to the occultation retrievals,
especially in terms of the narrowness of the \ctwohtwo\ mixing-ratio
peak shown in Fig.~\ref{fig:vtwosolingmix} and the \ctwohfour\ abundance
in the microbar region.  Similarly, models that take into account
seasonal variation in solar insolation \citep[based
  on][]{Moses05great} also provide a better fit than the 1-D
steady-state models shown in the figures here.

The mismatch between the model and data profiles may also, in theory,
result from an incorrect attribution of absorption to \ctwohfour,
which would increase its density at the expense of other species.  At
the resolution of the Voyager UVS, and once the absorption has
progressed to such a large extent that most of the diagnostic
wavelength range is completely absorbed, the similarity in the
photoabsorption cross sections of the various hydrocarbons considered
here makes this a possibility.  Retrieval techniques will always try
to find the best statistical match to the data, and the retrieved
\ctwohfour\ profile presented here represents that best match.  Close
examination of numerous individual model and data spectra (not shown)
does reveal that the \ctwohfour\ cross section is favored.  However,
as demonstrated by the comparison of the model light curves in
Fig.~\ref{fig:vtwosolinglc} to the data at the longest wavelengths,
very different density profiles (see Fig.~\ref{fig:vtwosolingret}) can
yield similar light curves.  Therefore, it may be the retrieved
profiles for \ctwohtwo\ and \ctwohfour, considered together, represent
the total opacity of both species rather than the individual opacity,
and some trade between \ctwohtwo\ and \ctwohfour\ is possible in the
retrieved density profiles.  Nevertheless, the fact that similar
data-model mismatches result from analysis of the higher resolution
Cassini UVIS data indicates that this issue is unlikely to be the only
factor involved and that some photochemistry-related problem still
exists.

An additional possibility is that other hydrocarbon species that were
not included in the data retrievals contribute to the opacity at the
wavelengths of the Voyager UVS.  To test this possibility, we modeled
light curves using our ``best-fit hydrocarbon'' model for this
occultation including the full range of species listed in
Table~\ref{tab:photocs}.  These added species contributed less than
2\% over the entire range of wavelengths longer than H~Lyman~$\alpha$,
and most of that was confined to only a few channels of the Voyager
UVS.  These model light curves are indistinguishable from those in
Fig.~\ref{fig:vtwosolinglc} at the scale of the figures and are not
shown.  This test indicates that the hydrocarbon species retrieved
from the Voyager occultations are by far the dominant species, to the
extent the photochemical models are valid, and that inclusion of
additional species in the retrievals is not warranted.  Species not
considered in Table~\ref{tab:photocs} (e.g., C$_{\rm 4}$H$_{\rm 10}$
and C$_{\rm 6}$H$_{\rm 6}$) are of sufficiently low abundance at
relevant altitudes in the models that they were not considered.  We
note that $\cfourhtwo$ {\em was} included in the data retrievals but
resulted in a statistically insignificant profile in all cases and
thus is not shown in the figures.  The contribution of $\cfourhtwo$
is, however, included in the light curves based on the photochemical
models.  On the other hand, if ion chemistry, which is not considered
in our photochemical models, contributes to the high-altitude
production of complex hydrocarbons, as it does on Titan
\citep[e.g.,][]{Waite07}, then further models that include ion
chemistry should be developed to better address the possible abundance
of complex neutrals in the occultation regions.  Note also that the
Cassini near-infrared stellar occultations show evidence for
high-altitude hazes \citep{Nicholson06,Bellucci09,Kim12}, which could
potentially be affecting the Voyager UVS light curves.

Turning to the final hydrocarbon retrieved from the Voyager~2 solar
ingress, we note that the inferred decrease in the \ctwohsix\ density
and mixing ratio in the lower portion of the retrievals is unlikely to
be real because \ctwohsix\ is a relatively stable molecule in this
altitude region.  The models predict that it should diffuse smoothly
into the lower atmosphere, and a reduction in the \ctwohsix\ density
with decreasing altitude is not expected based on what we know about
the hydrocarbon chemistry and atmospheric transport.  It is possible
that strong winds are affecting the profile, but it is more likely
that the mathematical inversion process used to retrieve this profile
has difficulty in correctly attributing the source of the absorption
in this region.  This conclusion is supported by the fact that
\ctwohsix\ absorption can be only be distinguished from absorption by
other hydrocarbons only over a very narrow range of wavelengths.
Within the uncertainties in the retrievals, however, the data profiles
are mostly consistent with the model profile for the ``best-fit
hydrocarbon'' model.

\subsection{Voyager 1 Solar Egress}
\label{sec:vonesoleg}

\subsubsection{Data Retrievals}

The Voyager~1 solar egress occultation, while providing similar
information to that of the Voyager~2 solar ingress, was much different
for a number of reasons.  A mid-occultation gain change (already noted
above) combined with large excursions in the pointing complicate the
generation of the transmission spectra.  However, because this
occultation is the primary origin of the higher thermospheric
temperature noted by \citet{Broadfoot81sat}, we have endeavored to
handle it as carefully as possible.

\begin{figure}[p]
    \includegraphics[width=6.5in]{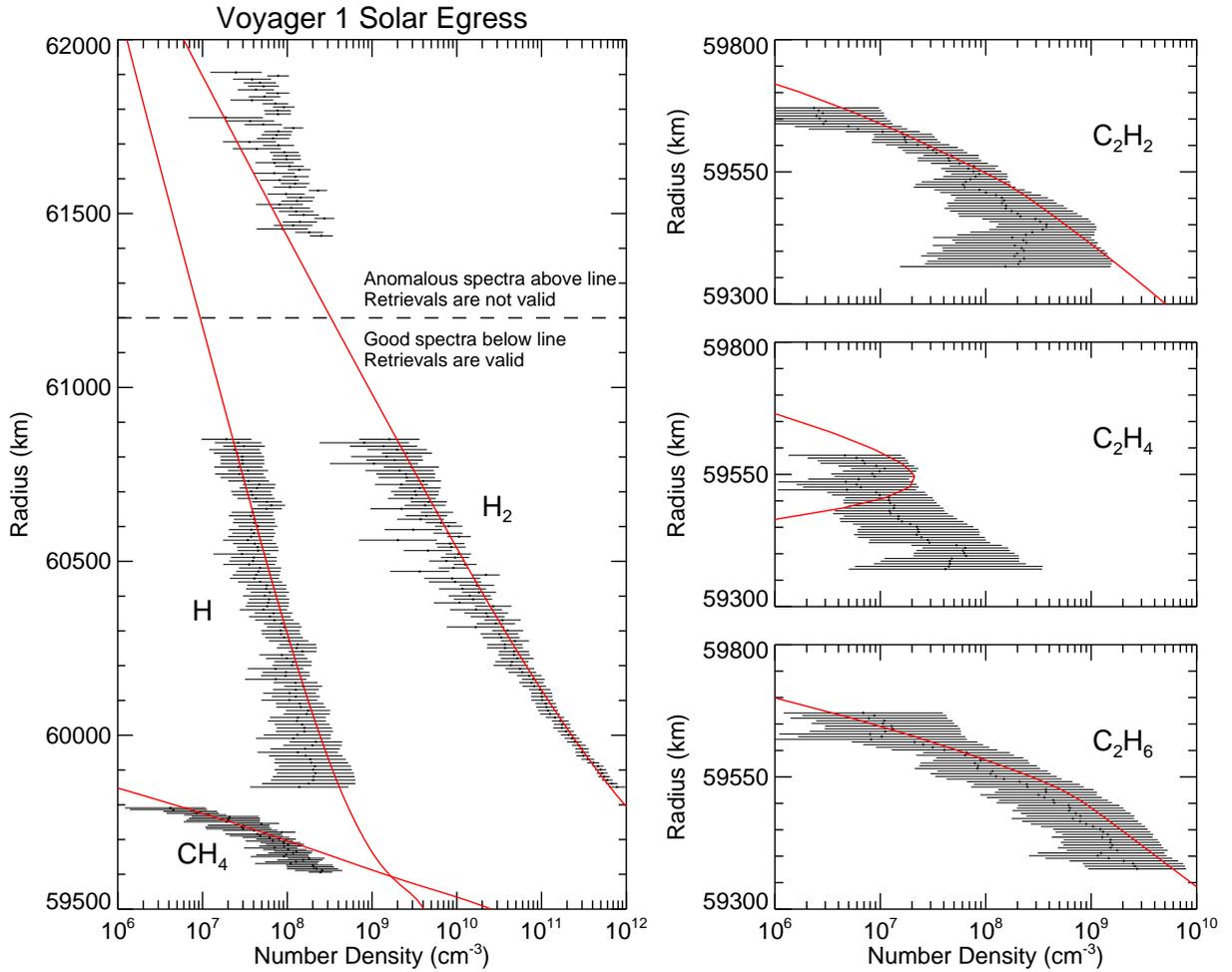}
    \caption{The density profiles from our forward model (red curve)
    for the Voyager~1 solar egress occultation at $-27\deg$ latitude
    are compared with the densities retrieved from our data inversions
    (black symbols, with associated one-sigma uncertainties).  The
    left figure is for \htwo, H, and \chfour, while the right panels
    are for \ctwohtwo\ (top), \ctwohfour\ (middle), and \ctwohsix\
    (bottom).  Only every other point is shown for the \htwo\ and H
    profiles for clarity.  See text for a discussion of the anomalous
    spectra and the implications of the retrieved \htwo\ profiles
    above and below the dashed line.}
    \label{fig:vonesolegret}
\end{figure}

\begin{figure}[p]
    \includegraphics[width=6.5in]{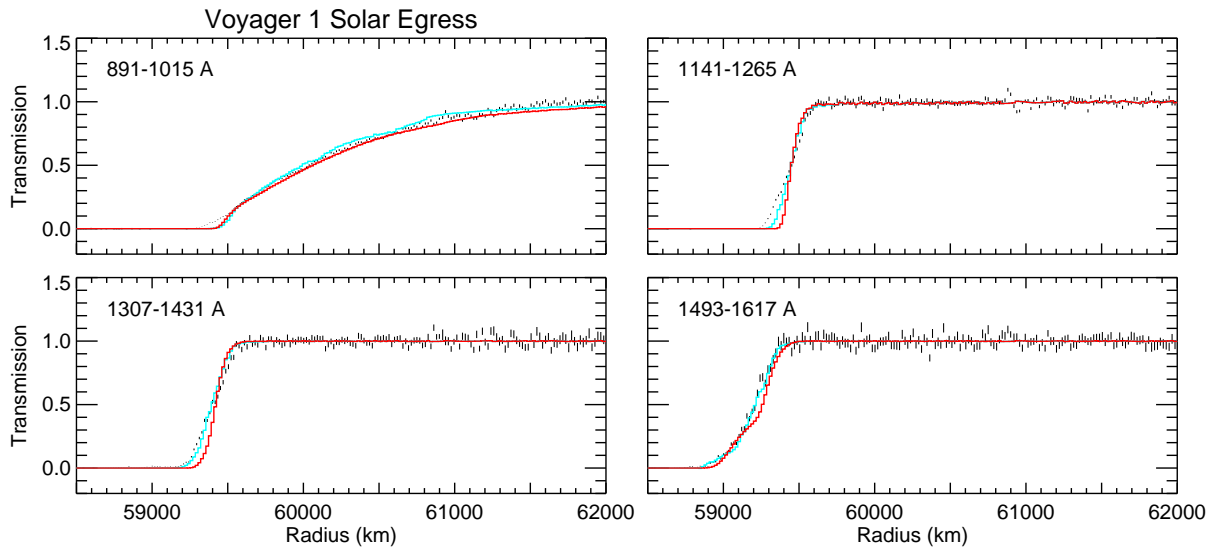}
    \caption{Light curves from our reanalysis of the Voyager~1 UVS
    solar egress occultation compared with synthetic light curves
    generated from our photochemical model (red) and from the
    retrieved density profiles (light blue).  The data are shown as
    black bars representing the one-sigma levels at each point.  These
    curves are averages of several channels and span the wavelength
    ranges indicated in each panel (including limit-cycle-induced
    shifts).  Only every sixth point is shown for clarity.  The top
    left panel is dominated by \htwo\ absorption at higher radii and
    \chfour\ at lower radii, while the top right panel is dominated by
    \chfour\ absorption at all altitudes.  The bottom two panels show
    aborption largely caused by complex hydrocarbons.}
    \label{fig:vonesoleglc}
\end{figure}

Figures~\ref{fig:vonesolegret} and~\ref{fig:vonesoleglc} show the
retrieved density profiles and data and synthetic light curve
comparisons, respectively, for the Voyager~1 solar egress occultation.
As with the Voyager~2 solar ingress, only every other point is shown
for \htwo\ and H for clarity, the hydrocarbons profiles are retrieved
using a tighter smoothing constraint than was used for the \htwo\ and
H profiles, and the data light curves and uncertainties are averages
over the specified wavelength ranges.  The same general features in
the retrieved density profiles as were noted for the Voyager~2 solar
ingress retrievals are also present in the Voyager~1 solar egress case
below the dashed horizontal line at 61200~km.  Above this line,
however, the \htwo\ profile for the Voyager~1 solar egress occultation
is different from that for the Voyager~2 solar ingress.

\clearpage

\clearpage

Clearly, the scale height of the Voyager~1 solar egress \htwo\ profile
above 61200~km changes, which leads us to several possible scenarios:
(1)~the change in scale height may reflect a real change in
temperature; (2)~\htwo\ opacity has been mistaken for H opacity
(notice the similarity to the H profile scale height); or (3)~the data
are bad.  Scenario~2 can be ruled out because the cross sections for
the \htwo\ and H continua are different enough to tell the two apart.
Furthermore, because the cross sections for \htwo\ are larger than
those for H at the wavelengths from which these densities are
determined, the implied H densities at these altitudes would be even
higher (roughly a factor of three) than the \htwo\ densities shown,
leaving a significant discontinuity between the H densities above
61200~km and those measured below and shown in
Fig.~\ref{fig:vonesolegret}.  The spectral fits at lower altitudes
cannot support such a high H density, so a large discontinuity would
be unavoidable and difficult to explain on a physical basis.

The first scenario~---~a temperature change~---~cannot be completely
ruled out, and it is most likely this enhanced scale height at higher
altitudes that led to the 850~K temperature determined in the earlier
analysis \citep{Broadfoot81sat}. In fact, a simple fit of an
exponential curve to these highest altitude density data yields a
temperature of 798$\pm$57~K.  However, we consider it unlikely and
favor the third scenario: bad data.  The lack of a similar change in
the higher-quality Voyager~2 solar ingress \htwo\ profile supports
this conclusion, but there is further evidence that makes the scenario
more certain.  As noted by \citet{Vervack04} for the Voyager~1 UVS
egress occultation at Titan (which occurred before the Saturn
occultations), the Voyager~1 UVS exhibited peculiar behavior at the
shorter wavelengths, and that strongly appears to be the case here.
Many of the channels in the \htwo\ continuum region between 600 and
850~\AA\ for Voyager~1 are systematically low relative to neighboring
channels when compared to the nominal solar spectrum shape generally
observed by the Voyager~1 UVS.  The net effect of these low signals is
to force the retrieval to higher densities to find the best overall
match to the data.

Additional evidence for anomalous behavior is that absorption at the
shortest wavelengths is complete long before absorption at wavelengths
greater than 850~\AA\ is observed, leading to the large gap in the
\htwo\ density profile in Fig.~\ref{fig:vonesolegret}.  Line-of-sight
column densities over this range would all be equal because there is
no changing absorption signature.  This clearly is not the case in the
Voyager~2 solar ingress occultation, and it makes no physical sense in
the Voyager~1 solar egress occultation because it implies decreasing
\htwo\ densities with decreasing altitude.  We have therefore chosen
only to show the retrieved \htwo\ densities above and below this
problem region: above to illustrate the likely origin of the
\citet{Broadfoot81sat} 850~K temperature from bad spectra, and below
where the retrieved densities are derived from valid spectra.  Note
that from this point on, we will not discuss the
\citet{Broadfoot81sat} results further.

We note that \citet{Smith83} concluded that pointing issues and the
gain change led to the higher temperatures in the initial analyses,
but they did not elaborate.  The $\Delta W$ motion of the spacecraft
at this point in the Voyager~1 solar egress occultation did place the
Sun close to the edge of the slit, therefore leading to a smaller
input signal to the UVS simply due to the mechanical collimator.  This
fact, coupled with the lower HVL~2 gain of the UVS at the time of
these observations, would have led to very few counts in the region of
the solar spectrum where the anomalous behavior is observed and might
possibly mimic the problems seen in the solar spectra.  However, the
pointing and gain change affect all channels of the spectrum, and
comparisons of HVL~2 solar spectra from regular solar calibration
observations when the Sun is at the same point in the slit do not show
the same spectral shape as observed in the Voyager~1 solar egress
before absorption set in (i.e., in the reference spectra range).
Therefore, we believe the answer to the anomalous behavior lies
elsewhere, at least in part.

A possible explanation is that the observed behavior is connected to
part of the UVS detector.  The detectors were fabricated in sections
of sixteen channels, and the problem channels appear to be confined to
one or two adjoining sections.  Whether it is a possible problem with
the detector itself or the read-out electronics associated with these
sections, we cannot say, but there is enough evidence to suggest that
this region of the UVS detector was unreliable during the Saturn and
Titan encounters, at least at the high signal levels of a solar
occultation.  Luckily, Saturn was the last planetary encounter for
Voyager~1.

The net result of this problem with the spectra at short wavelengths
is that the usable portion of the Voyager~1 solar egress retrievals
does not probe to as large radial distances as the Voyager~2 solar
ingress.  Therefore, no panels corresponding to the upper left and
upper right panels of Figure~\ref{fig:vtwosolinglc} are shown in
Figure~\ref{fig:vonesoleglc}.  However, the light curves and
retrievals corresponding to lower radial distances are of good quality
and provide the second-best set of Voyager UVS occultation data from
Saturn.

\subsubsection{Photochemical Modeling}

The Voyager~1 solar egress occultation differs from the Voyager~2
solar ingress occultation mainly in that absorption due to methane and
the other hydrocarbons sets in at a lower altitude relative to the
\htwo\ absorption, suggesting a lower-altitude homopause level and
less vigorous atmospheric mixing.  We developed several forward models
for the Voyager~1 solar egress occultation; the temperature profile
from our best-fit model can be found in the journal's supplementary
material and section~\ref{sec:stellardata}.  Although we still require
a relatively warm lower stratosphere to provide enough \htwo\ density
in the microbar region to compare with the occultation light curves,
we find that our models fit the density retrievals better when the
mesopause is located at relatively low altitudes (high pressures) as
compared with the Voyager~2 solar ingress occultation.  This solution
makes sense physically, as the hydrocarbons appear to be diffusing out
at higher pressures at the latitude of this occultation ($-27\deg$),
leaving no efficient cooling agents at higher altitudes.

Our favored model for the Voyager~1 solar egress occultation has an eddy
diffusion coefficient profile that fits the following expression:
$$
%\halign to \hsize{$#$\hfill & $#$\hfill & $#$\hfill\quad \cr
\halign to \hsize{$#$\hfill & $#$\hfill & $#$\hfill\quad & $#$\hfill\quad \cr
K_{zz} & \ = \ & 1.0\scinot7.\, (2.0\scinot-4./p)^{0.5} & \ \ {\rm For}\  p < 2.0\scinot-4.\ \hbox{mbar} \cr
%\ &   &  \ \ \ {\rm For}\  p < 2.0\scinot-4.\ \hbox{mbar} \cr
\ & \ = \ & 1.0\scinot7.\, (2.0\scinot-4./p)^{0.78} & \ \ {\rm For}\ 2.0\scinot-4. \le p < 0.5\ \hbox{mbar} \cr
%\ &   &  \ \ \ {\rm For}\ 2.0\scinot-4. \le p < 0.5\ \hbox{mbar} \cr
\ & \ = \ & 2.237\scinot4.\, (0.5/p)^{0.68} & \ \ {\rm For}\ 0.5 \le p < 150\ \hbox{mbar} \cr
%\ &   &  \ \ \ {\rm For}\ 0.5 \le p < 150\ \hbox{mbar} \cr
\ & \ = \ & 5.0\scinot2. & \ \ {\rm For}\ 150 \le p < 630\ \hbox{mbar} \cr
%\ &   &  \ \ \ {\rm For}\ 150 \le p < 630\ \hbox{mbar} \cr
\ & \ = \ & 2.5\scinot4.  & \ \ {\rm For}\ p > 630\ \hbox{mbar} \cr}
%\ &   &  \ \ \ {\rm For}\ p > 630\ \hbox{mbar} \cr}
$$
for $K_{zz}$ in cm$^2$ s$^{-1}$ and $p$ in mbar.  The other model 
assumptions are similar to those from our ``standard'' model for the 
Voyager~2 solar ingress occultation.

Figure~\ref{fig:vonesolegret} shows how the density profiles from this
forward model compare with our retrievals, and
Fig.~\ref{fig:vonesoleglc} shows how synthetic light curves generated
from the model results compare with the observed light curves.  In
both instances, the high-altitude behavior is well reproduced, but
problems exist at lower altitudes.  Although the observed half-light
point in the light curves at methane-sensitive wavelengths is at
roughly the same radius in this occultation ($-27\deg$ latitude) as
compared with the Voyager~2 solar ingress occultation
(29\deg\ latitude), the gravity field and atmospheric structure are
very different at these two latitudes, and the same radius translates
to a different altitude and pressure.  The greater centrifugal
acceleration and the stronger zonal winds at $-27\deg$ as compared to
29\deg\ latitude act such that the 1-bar radius is at least 150~km
larger at $-27\deg$ latitude than at 29\deg\ latitude
\citep[e.g.,][]{Lindal85,SanchezLavega00}.  From either an
extrapolation of the \htwo\ densities or from our forward modeling, we
infer the half-light point for methane to be at a higher pressure in
the $-27\deg$ occultation than the 29\deg\ occultation, implying less
vigorous atmospheric mixing at $-27\deg$ latitude at the time of the
occultation.  For the Voyager~1 solar egress model at $-27\deg$
latitude, the methane half-light point is 59452~km (681~km above the
1-bar level) or a pressure of $\sim$1.2$\scinot-3.$ mbar, at which
point $K_{zz}$ $\approx$ 2.5$\scinot6.$ cm$^2$ s$^{-1}$.  The
half-light point for methane is therefore inferred to be at a pressure
roughly 40-to-50 times larger at $-27\deg$ latitude than at
29\deg\ latitude (i.e., about one and a half scale heights deeper), at
which point the eddy diffusion coefficient is 30-to-40 times smaller.

\begin{figure}[p]
    \includegraphics[scale=0.75]{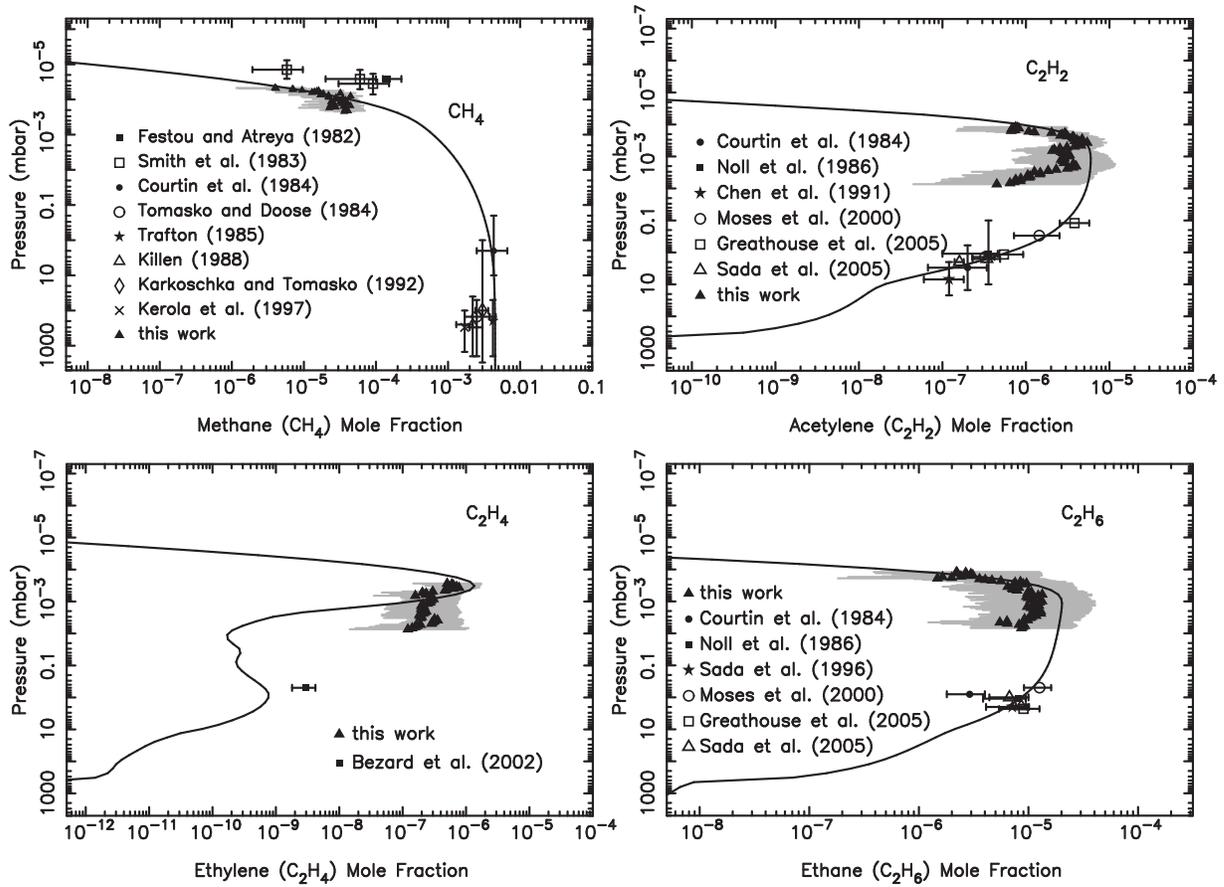}
    \caption{The mixing ratios of \chfour\ (top left), \ctwohtwo\ (top
    right), \ctwohfour\ (bottom left), and \ctwohsix\ (bottom right)
    in our forward model (red) of the Voyager~1 solar egress
    occultation, as compared with our retrievals (solid triangles) and
    various other pre-Cassini observations (as labeled).  The pressure
    scale for the UVS retrievals plotted here is taken from the
    radius-vs-pressure grid from the forward model and is very model
    dependent.  Note that the inferred \chfour\ homopause in this
    occultation is at higher pressures than that inferred from the
    Voyager~2 solar ingress occultation at 29\deg\
    latitude.}
    \label{fig:vonesolegmix}
\end{figure}

The model-data comparisons are not good in the lower portion of the
occultations, especially at hydrocarbon-sensitive wavelengths.  Unlike
the situation for the 29\deg\ latitude modeling, our $-27\deg$
latitude model produces too much hydrocarbon absorption (except for
\ctwohfour) at low altitudes compared to what can be allowed from the
observations.  The methane concentration, in particular, is too great.
As can be seen from Fig~\ref{fig:vonesolegmix}, the inferred \chfour\ mixing
ratio drops with altitude from a peak near 10$^{-4}$ mbar.  This
behavior is not possible with diffusion alone; if real, the observed
structure must be caused by wave activity or other winds with vertical
shear.  In contrast, the \ctwohfour\ concentration is greatly
underpredicted by the models at low altitudes.  As discussed earlier,
the similar nature of this problem for all the occultation model-data
comparisons here and with the Cassini UVIS analysis
\citep[see][]{Shemansky12,Nagy09} suggests a shortcoming in the
photochemical model.  In particular, the model seems to be missing an
effective pathway for converting \ctwohtwo\ to \ctwohfour\ in the
microbar region on Saturn.

\clearpage

\subsection{Voyager 1 and 2 Stellar Occultations}
\label{sec:stellardata}

\subsubsection{Data Retrievals}

The Voyager~1 and~2 stellar occultation data are of lower
signal-to-noise than the solar occultation data owing to the vastly
different intensity of the stars versus the Sun.  The star used for
the Voyager~2 stellar occultations was $\delta$ Scorpii (spectral type
B0.2IV with $m_B$=2.2 and $m_V$=2.3); the star $\iota$ Herculis
(spectral type B3IV with $m_B$=3.6 and $m_V$=3.7) was used for the
Voyager~1 stellar egress occultation.  Nevertheless, by averaging
channels across wavelengths, we are able to retrieve \htwo\ and
\chfour\ density profiles for each of the three stellar occultations.
However, meaningful retrievals of C$_2$H$_x$ hydrocarbon densities
have not been obtained (the resulting uncertainties are too large)
although the light curves at those wavelengths can still provide some
level of check on the heavier hydrocarbons in models if they are
heavily averaged in both altitude and wavelength.  Owing to absorption
by interstellar H below 912~\AA\ in the stellar spectra, H could not
be retrieved.

In order to carry out the retrievals for these occultations, it is
necessary to average the data in wavelength.  For the Voyager~2
stellar ingress and egress, the data are averaged in groups of three
adjacent wavelength channels; however, for the Voyager~1 stellar
egress the averaging has to be done over ten adjacent channels owing
to the very weak intensity of the star.  No averaging is done as a
function of altitude for any occultation.

An additional consideration for the Voyager~2 stellar ingress
occultation is that it suffered from contamination by the rings.  This
contamination is evident in the data as ``bite-outs'' (localized dips)
in the signal as a function of radius.  Because these ``bite-outs''
affect every wavelength, it is easy to distinguish these occurrences
from actual atmospheric absorption.  The regions affected are isolated
and removed from the data and fortuitously do not significantly affect
either the region of absorption or the range of reference spectra that
are needed.  Only one small gap in the \htwo\ absorption region is
evident in the retrievals.

\begin{figure}[p]
    \includegraphics[width=6.5in]{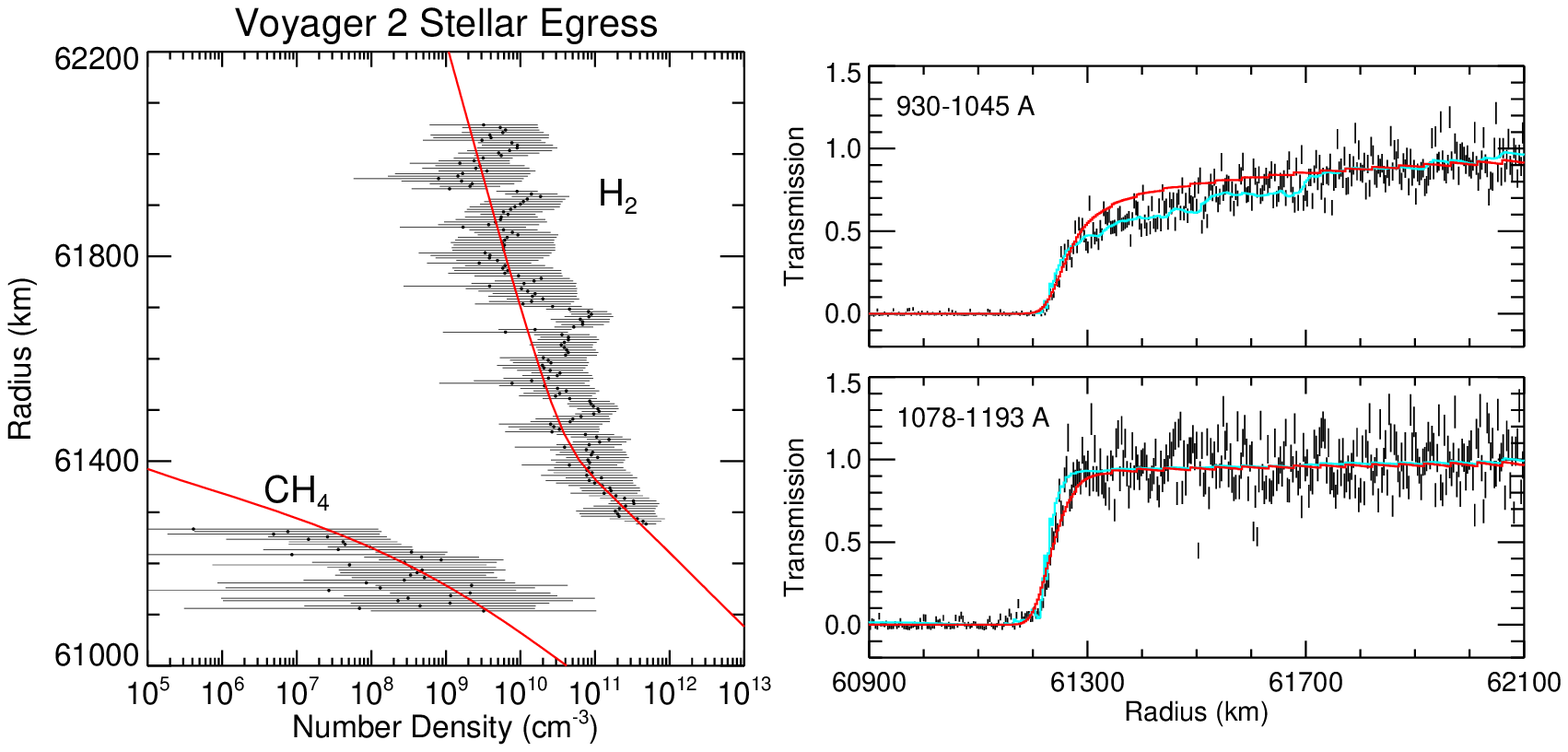}
    \caption{The density profiles (left panel) from our forward model
    (red curve) for the Voyager~2 stellar egress occultation at
    3.8\deg\ latitude are compared with the densities retrieved from
    our data inversions (black symbols, with associated one-sigma
    uncertainties).  Profiles for \htwo\ and \chfour\ are shown.
    Light curves (right panels) for this occultation are compared with
    synthetic light curves generated from our photochemical model
    (red) and from the retrieved density profiles (light blue).  The
    data are shown as black bars representing the one-sigma levels at
    each point.  These curves are averages of several channels and
    span the wavelength ranges indicated in each panel (including
    limit-cycle-induced shifts).  Both panels show absorption due to
    \htwo\ and \chfour; \htwo\ dominates in the top panel and \chfour\
    in the bottom.}
    \label{fig:vtwosteleg}
\end{figure}

\begin{figure}[p]
    \includegraphics[width=6.5in]{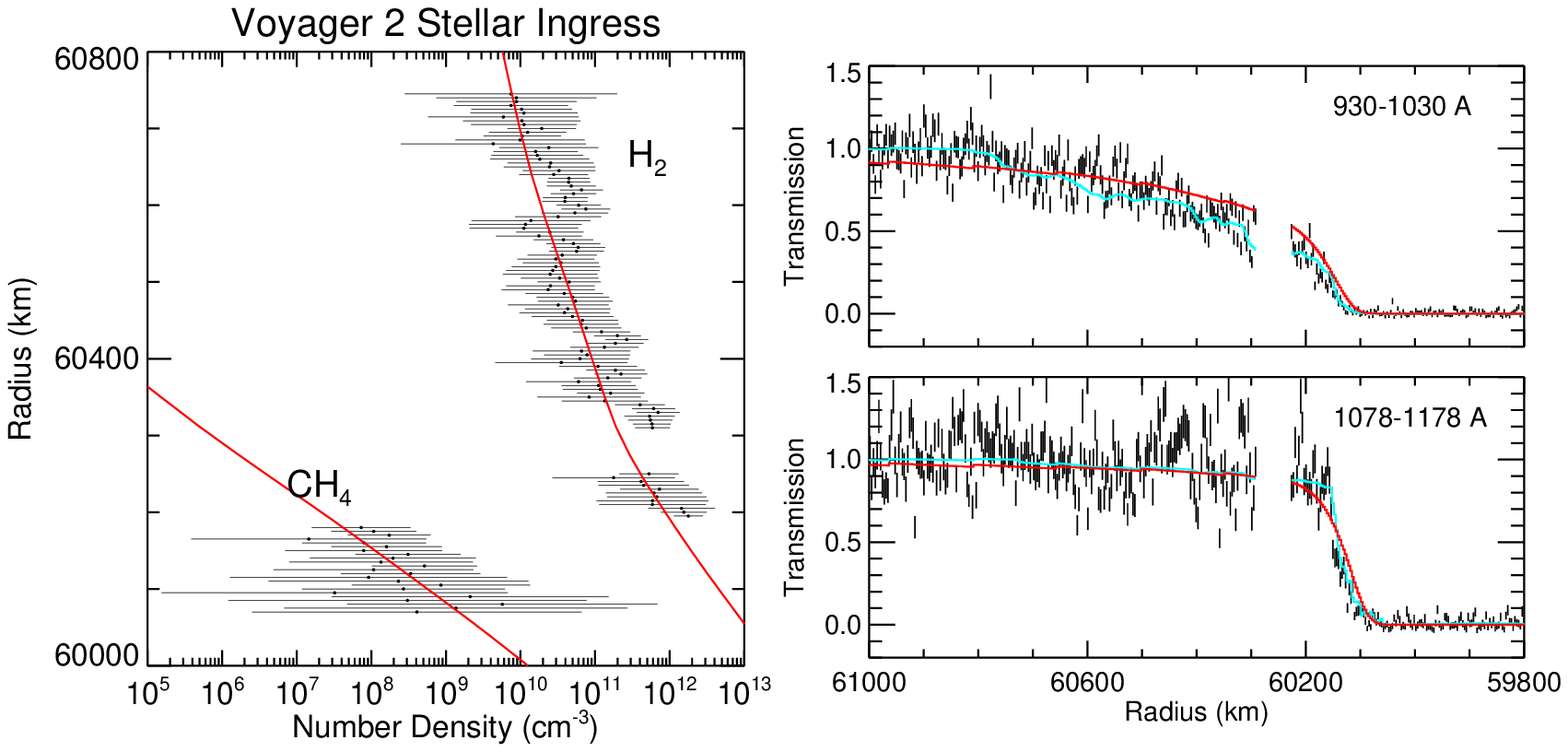}
    \caption{The density profiles (left panel) from our forward model
    (red curve) for the Voyager~2 stellar ingress occultation at
    $-21.5\deg$ latitude are compared with the densities retrieved
    from our data inversions (black symbols, with associated one-sigma
    uncertainties).  Profiles for \htwo\ and \chfour\ are shown.
    Light curves (right panels) for this occultation are compared with
    synthetic light curves generated from our photochemical model
    (red) and from the retrieved density profiles (light blue).  The
    data are shown as black bars representing the one-sigma levels at
    each point.  These curves are averages of several channels and
    span the wavelength ranges indicated in each panel (including
    limit-cycle-induced shifts).  Both panels show absorption due to
    \htwo\ and \chfour; \htwo\ dominates in the top panel and \chfour\
    in the bottom.  The gap is a region where spectra had to be
    eliminated owing to interference from Saturn's rings.}
    \label{fig:vtwosteling}
\end{figure}

\begin{figure}[p]
    \includegraphics[width=6.5in]{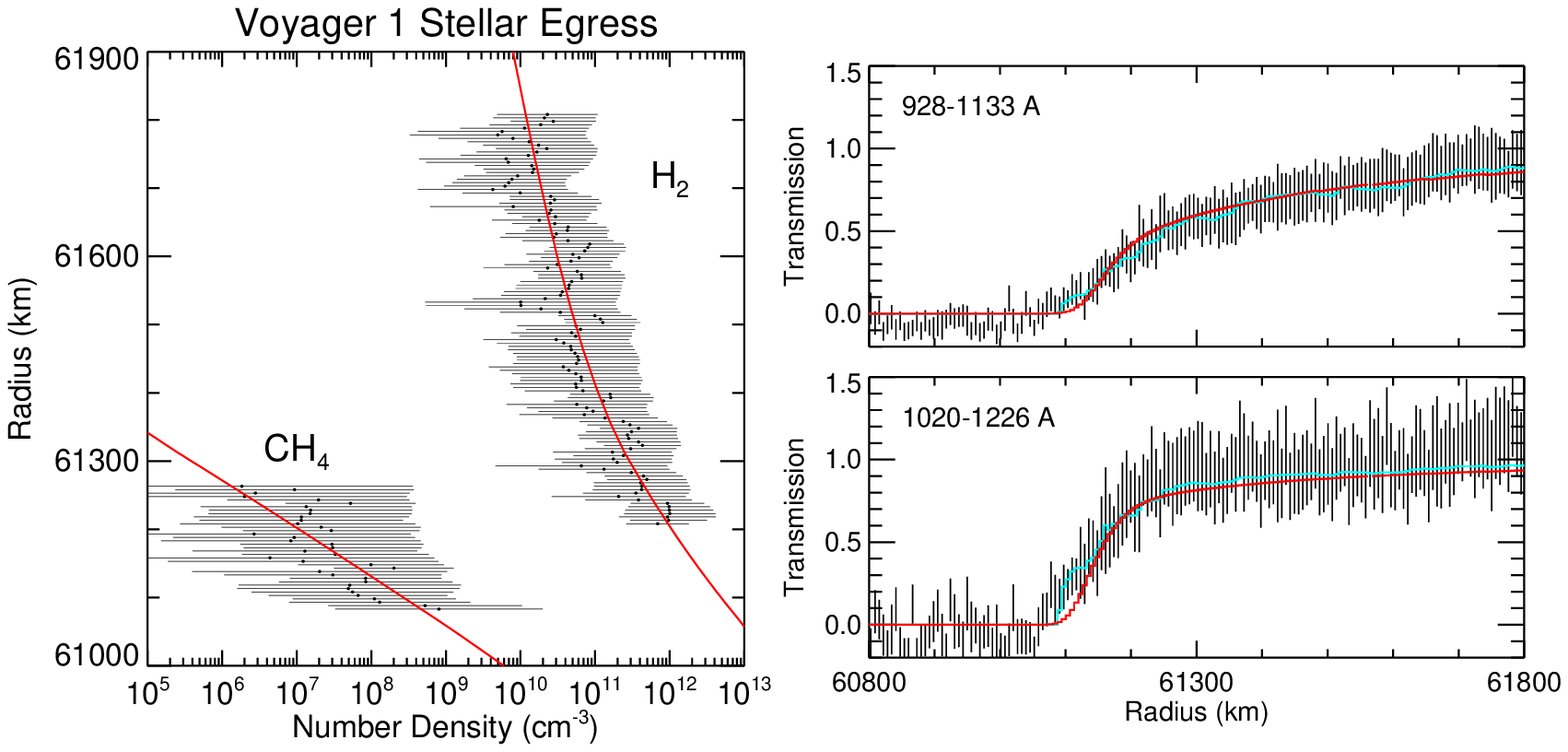}
    \caption{The density profiles (left panel) from our forward model
    (red curve) for the Voyager~1 stellar egress occultation at
    $-4.8\deg$ latitude are compared with the densities retrieved
    from our data inversions (black symbols, with associated one-sigma
    uncertainties).  Profiles for \htwo\ and \chfour\ are shown.
    Light curves (right panels) for this occultation are compared with
    synthetic light curves generated from our photochemical model
    (red) and from the retrieved density profiles (light blue).  The
    data are shown as black bars representing the one-sigma levels at
    each point.  These curves are averages of several channels and
    span the wavelength ranges indicated in each panel (including
    limit-cycle-induced shifts).  Both panels show absorption due to
    \htwo\ and \chfour; \htwo\ dominates in the top panel and \chfour\
    in the bottom.}
    \label{fig:vonesteleg}
\end{figure}

Figures~\ref{fig:vtwosteleg}, \ref{fig:vtwosteling}, and
\ref{fig:vonesteleg} (left panels) show the \htwo\ and
\chfour\ profiles retrieved from the Voyager~2 stellar egress,
Voyager~2 stellar ingress, and Voyager~1 stellar egress occultations,
respectively.  Those figures also show comparisons of the data light
curves to light curves synthesized using the retrieved density
profiles.  The two panels on the right of each figure show wavelength
regions dominated by \htwo\ (top right panels) and \chfour\ (bottom
right panels) absorption.  Despite the weaker nature of the stars, the
retrievals for these occultations have yielded good measurements of
the \htwo\ and \chfour\ density profiles.

\clearpage

\clearpage

\clearpage

We note that this is the first time the Voyager~2 stellar ingress and
Voyager~1 stellar egress have been analyzed.  In the case of the
former, the interference by the rings made the processing and
retrieval complicated enough that it was never examined in detail
despite its good quality otherwise.  In the case of the latter, the
weak star made for a very limited analysis.  However, to provide
greater latitudinal coverage, we have extracted as much information as
possible from this occultation.

\subsubsection{Photochemical Modeling}

\begin{figure}[p]
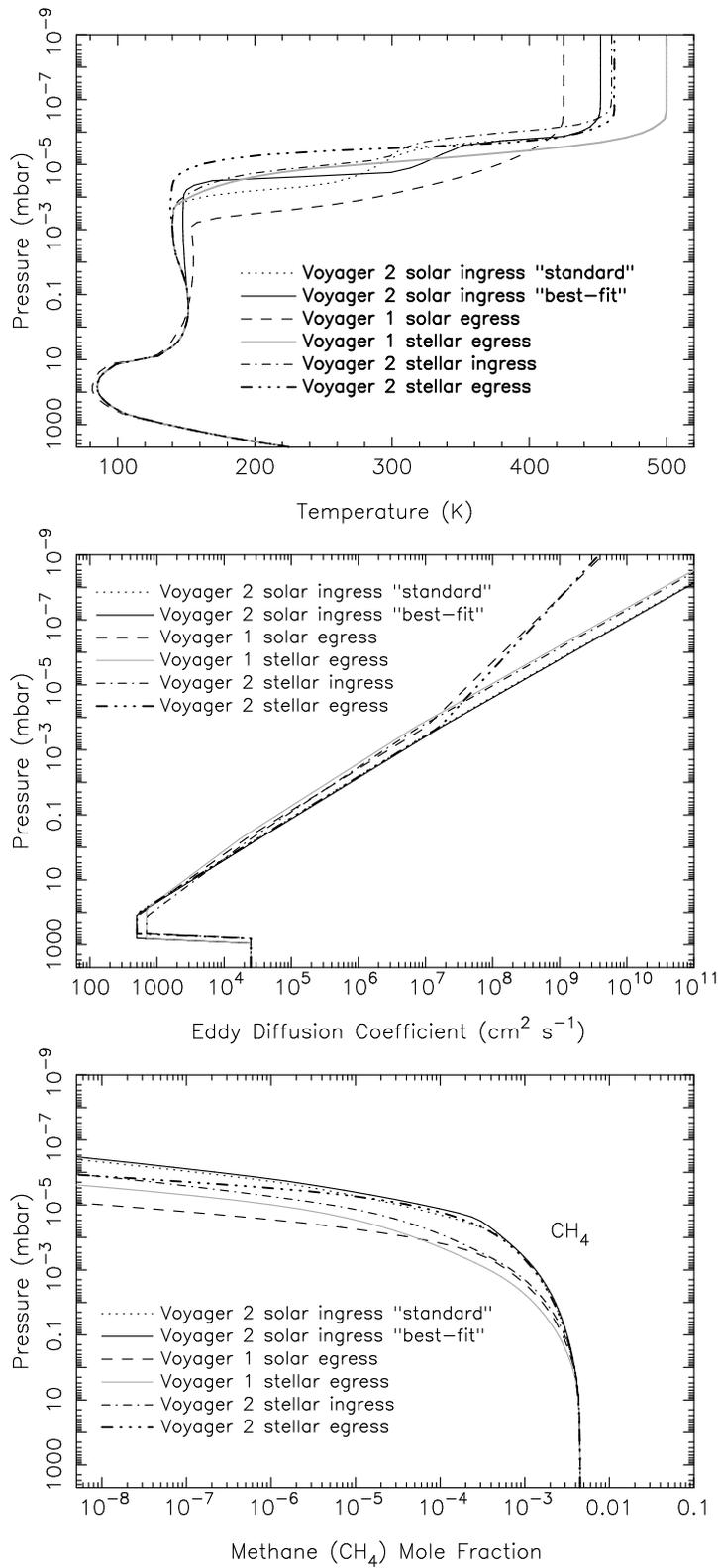

    \begin{center}
        \mbox{
        \begin{minipage}[t]{.7\hsize}
            \includegraphics[scale=0.4,angle=-90]{figures/modtempall.ps}
        \end{minipage}}
    \end{center}
    \begin{center}
        \mbox{
        \begin{minipage}[t]{.7\hsize}
            \includegraphics[scale=0.4,angle=-90]{figures/modeddyall.ps}
        \end{minipage}}
    \end{center}
    \begin{center}
        \mbox{
        \begin{minipage}[t]{.7\hsize}
            \includegraphics[scale=0.4,angle=-90]{figures/modch4all.ps}
        \end{minipage}}
    \end{center}
    \caption{The temperature profiles (top) and eddy diffusion coefficient
    profiles (middle) assumed for our forward models.  The resulting 
    model \chfour\ profiles are shown in the bottom figure.}
    \label{fig:tempeddy}
\end{figure}

The retrieved \htwo\ and \chfour\ density profiles for each stellar
occultation are of high enough quality that we can generate forward
models to at least determine the methane homopause levels or the
values of $K_{zz}$ at relevant altitude or pressure levels.
Figures~\ref{fig:vtwosteleg}, \ref{fig:vtwosteling}, and
\ref{fig:vonesteleg} show how our forward models compare with the
density retrievals, as well as how synthetic light curves from these
models compare with the observations.  Our model assumptions about the
temperature and $K_{zz}$ profiles are shown in Fig.~\ref{fig:tempeddy}.

\clearpage

The models do an acceptable job of reproducing both the retrieved
densities and the light curves.  The signal-to-noise in the stellar
occultations is low enough that the slopes of the \chfour\ profiles
--- and hence the $K_{zz}$ profiles --- are hard to uniquely
constrain, but we can at least provide some constraints on the heights
to which the methane is carried.  For the Voyager~1 stellar egress
occultation at $-4.8\deg$ latitude, the half-light point for the
methane-sensitive wavelengths is at 61142~km (929~km above the 1-bar
radius), corresponding to a model pressure level of
$\sim$6.0$\scinot-5.$, at which point $K_{zz}$ $\approx$
2.0$\scinot7.$ cm$^2$ s$^{-1}$.  For the Voyager~2 stellar ingress
occultation at $-21.5\deg$ latitude, the half-light point for the
methane-sensitive wavelengths is at 60124~km (864~km above the 1-bar
radius), corresponding to a model pressure level of
$\sim$6.6$\scinot-5.$, at which point $K_{zz}$ $\approx$
2.1$\scinot7.$ cm$^2$ s$^{-1}$.  For the Voyager~2 stellar egress
occultation at 3.8\deg\ latitude, the half-light point for the
methane-sensitive wavelengths is 61232~km (1000~km above the 1-bar
radius), corresponding to a model pressure level of
$\sim$1.8$\scinot-5.$ mbar, at which point $K_{zz}$ $\approx$
4.7$\scinot7.$ cm$^2$ s$^{-1}$.  Those heights are best illustrated by
the mixing-ratio plot in Fig.~\ref{fig:tempeddy}, which shows a
significant variation between the different occultations.  A
convenient measure of the differences of the strengths of eddy mixing
between the different latitudes is the pressure level at which the
methane mole fraction drops to 5$\scinot-5.$ in the models, and the
$K_{zz}$ value corresponding to that pressure level (see
Table~\ref{tab:chfour}).  That mole fraction value is sampled within
the occultation regions for all the different models.  Note from
Fig.~\ref{fig:tempeddy} and Table~\ref{tab:chfour} that atmospheric
mixing was very vigorous at 29\deg\ and 3.8\deg\ latitude at the time
of the Voyager~2 solar ingress and stellar egress occultations and
much less vigorous at $-27\deg$ and $-4.8\deg$ latitude at the time of
the Voyager~1 solar egress and Voyager~1 stellar egress occultations.

\begin{table}[h]
\caption{Model Parameters at Level Where CH$_4$ Mole Fraction is 5$\scinot-5.$}
\begin{tabular}{lcccc}
\hline
\           & Planetocentric & Pressure  & $K_{zz}$          & Local Time$^{\rm a}$ \\
Occultation & Latitude       & (mbar)    & (cm$^2$ s$^{-1}$) & (average) \\
\hline 
Voyager 2 solar ingress        & $29\deg$ &        &                        & 8.245 \\
\ \ \ \ \ ``standard model''   &  & 1.3$\scinot-5.$ & 1.6$\scinot8.$        & \\
\ \ \ \ \ ``best-fit hydrocarbon'' &  & 8.9$\scinot-6.$ & 2.4$\scinot8.$    & \\
Voyager 2 stellar egress   & $3.8\deg$ & 1.1$\scinot-5.$ & 6.0$\scinot7.$   & 9.63 \\ 
Voyager 1 stellar egress   & $-4.8\deg$  & 1.0$\scinot-4.$ & 1.3$\scinot7.$ & 10.56 \\
Voyager 2 stellar ingress  & $-21.5\deg$ & 4.1$\scinot-5.$ & 3.1$\scinot7.$ & 4.65 \\ 
Voyager 1 solar egress     & $-27\deg$   & 1.1$\scinot-4.$ & 1.4$\scinot7.$ & 2.77 \\
\hline
\end{tabular}

$^{\rm a}$ Local time is defined using a Saturn rotational period of 10.76
           hours (i.e., ``noon'' is 5.38 and ``midnight'' is 10.76).  These are
           averages of the values in Table~\protect\ref{tab:geometry}.
\label{tab:chfour}
\label{lasttable}
\end{table}
\clearpage

\subsubsection{On the Different Temperatures Derived from the Voyager 2 Stellar Egress Occultation}
\label{sec:vtwofestou}

We are now in a position to address the possible reasons for the
vastly different temperatures derived for the Voyager 2 $\delta$ Sco
stellar egress occultation at 3.8\deg\ latitude, for which
\citet{Festou82} favored an exospheric temperature of
800$^{+150}_{-120}$ K, whereas \citet{Smith83} favored 420$\pm 30$ K.
Both analyses adopted a forward-modeling technique with a reference
level within the occultation region.  The data themselves help
constrain the \htwo\ density at this reference level, so both groups
have reasonable \htwo\ densities for at least part of the radial
profiles.  We argue that the differences between \citet{Smith83} and
\citet{Festou82} likely result from data-quality issues, from pitfalls
associated with attempts to derive temperatures from forward models
using noisy data, and from attempts to use stellar occultations rather
than solar occultations at \htwo\ continuum wavelengths to derive
exospheric temperatures.

\begin{figure}[p]
    \includegraphics[scale=0.59,angle=-90]{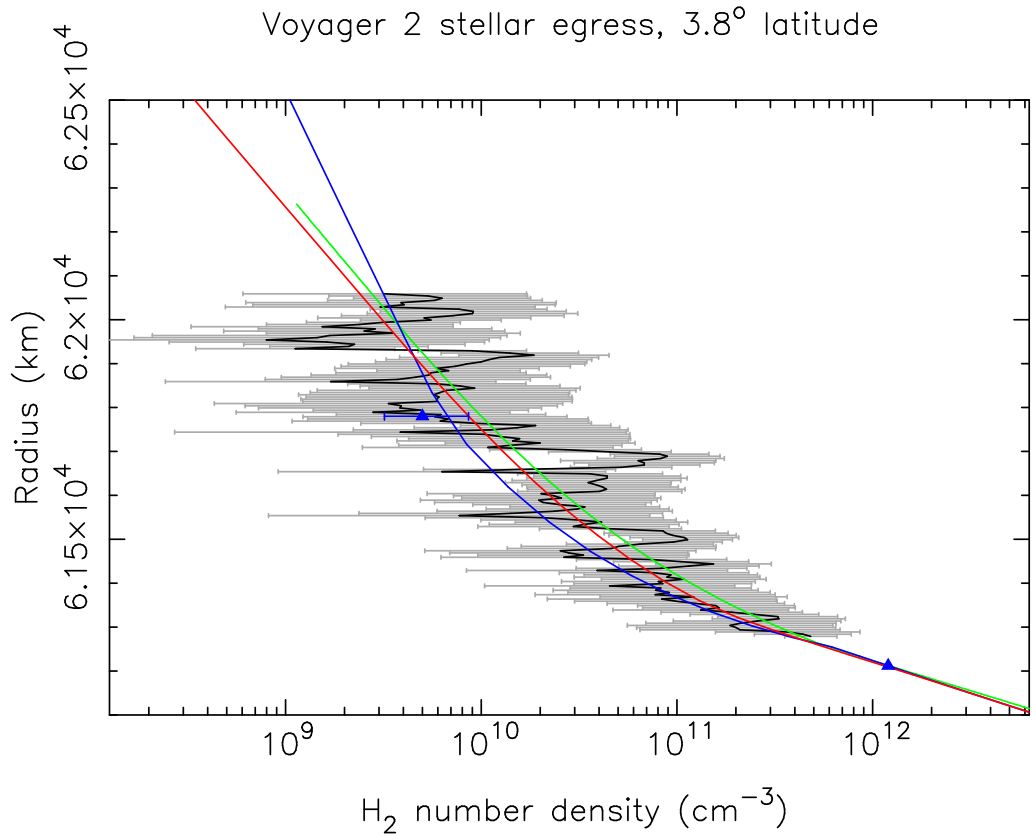}
    \caption{The \htwo\ number density as a function of radius for the
    Voyager 2 $\delta$ Sco stellar egress occultation at 3.8\deg\
    latitude from our data retrieval (black with associated gray error
    bars), compared with the model derived from the
    \protect\citet{Shemansky12} reanalysis of this occultation (green),
    and two of our own forward models that assume an exospheric
    temperature of 450~K (red) and 800~K (blue).  The blue triangles
    represent the \htwo\ density at two specific points discussed in
    the occultation analysis of \protect\citet{Festou82}.  Note the
    similarities in the \htwo\ densities inferred from all these
    techniques, despite the widely different assumed or derived
    temperature profiles.  This non-uniqueness is the likely cause of
    the different derived exospheric temperatures
    from \protect\citet{Festou82} and \protect\citet{Smith83} from the same
    occultation.  This comparison highlights the potential pitfalls of using
    forward models alone to determine temperatures from ultraviolet
    occultations.}
    \label{fig:festou}
\end{figure}

As an example of these problems and pitfalls,
Figure~\ref{fig:festou} shows two assumed forward-model temperature
profiles we developed for the $\delta$ Sco egress occultation region
probed by Voyager 2.  One profile (red curve) has a $\sim$450 K
exospheric temperature similar to that derived by \citet{Smith83},
whereas the other profile (blue curve) has an 800-K exospheric
temperature and a profile similar to that derived by \citet{Festou82}.
All other forward model parameters are the same between these two
models, and all models go through the \htwo\ = 1.2$\, \times\,
10^{12}$ cm$^{-3}$ point at 61212 km, as cited by \citet{Festou82}.
Also shown in Figure~\ref{fig:festou} are the model results from
the latest reanalysis of this occultation by \citet{Shemansky12}
(green curve), and the unsmoothed \htwo\ density structure from our
retrievals described earlier in section~\ref{sec:stellardata}.  The
triangles mark the radius-\htwo\ density points cited by
\citet{Festou82} in their text.  Note, in particular, that the derived
\htwo\ density of 5.0$^{+3.6}_{-1.8}\, \times\, 10^{9}$ cm$^{-3}$ at
61780 km compares very well with our \htwo\ retrieval at the same
altitude.

\clearpage

Although our retrieved \htwo\ profile provides the best mathematical
fit to the occultation light-curve data, all three model profiles with
their very different thermal structures provide a reasonable fit (to
within the noise level of the data) to the light curves at wavelengths
sensitive to \htwo\ absorption.  This exercise illustrates the
non-uniqueness of the forward modeling (see also Appendix A), and, in
particular, the difficulty in constraining temperatures from forward
models.  Forward modeling of noisy stellar occultation data is thus
limited in its ability to uniquely determine the exospheric
temperatures on Saturn, and care should be exercised in interpreting
such results.

\subsection{Voyager 1 Solar Ingress}
\label{sec:vonesoling}

Among the Voyager UVS occultations, the Voyager~1 solar ingress
occultation is highly unique owing to its location at $-83\deg$
latitude.  Unfortunately, the passage of Voyager under Saturn's
southern pole caused the spacecraft trajectory to change rapidly
during the occultation.  In order for the scan platform to keep the
UVS boresight oriented on the Sun, it had to slew at regular
intervals.  Data acquired during a slew are not good, and because
there is a finite settle time after a slew, data immediately after a
slew are also not viable.  The usable range of data therefore consists
of a number of small ranges of radii separated by gaps.

A further complication is that each time the scan platform executes a
slew, the limit cycle information, which is a relative measure of
position, resets.  Therefore, in order to generate the transmission
spectra, each segment of good data has to be tied to the adjacent
segment so that reference spectra acquired early in the occultation
can be placed in the same limit cycle frame as the absorption region
spectra.  To do this, the position of the H Lyman $\alpha$ line is used as
a reference point in each spectrum and the limit cycle for each
spectrum is tied to the known shift of the H Lyman $\alpha$ line relative
to its on-axis position.  In this manner, a consistent limit cycle
frame for the entire occultation can be constructed.

Although retrievals of \htwo\ and H are possible, unfortunately no
hydrocarbon retrievals can be carried out for two reasons.  First,
hydrocarbon absorption generally takes place very rapidly at the
homopause level owing to the nature of the hydrocarbon profiles as a
function of radius (i.e., fast fall-off near the homopause).  It just
happens that the majority of the range of hydrocarbon absorption falls
into one of the unusable gaps.  Second, once the H Lyman $\alpha$ line
is absorbed, the remaining segments of good data cannot be tied to the
others with high confidence.  Thus, even though there are some few
spectra for which retrievals may be possible, they cannot be placed in
the same frame as the others.

\begin{figure}[p]
    \includegraphics[width=6.5in]{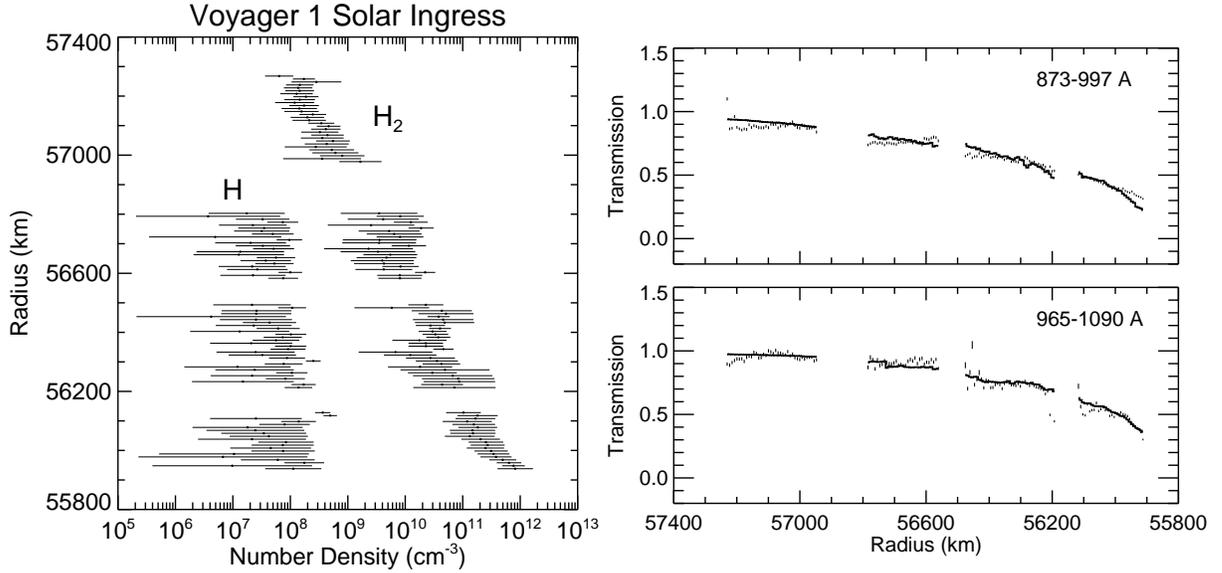}
    \caption{The density profiles (left panel) retrieved from our data
    inversion (black symbols, with associated one-sigma uncertainties)
    for the Voyager~1 solar ingress occultation at $-83\deg$ latitude.
    Profiles for \htwo\ and H are shown; only every other point in
    these retrieved density profiles is plotted for clarity.  Light
    curves (right panels) for this occultation are compared with
    synthetic light curves generated from the retrieved density
    profiles.  The data are shown as black bars representing the
    one-sigma levels at each point.  These curves are averages of
    several channels and span the wavelength ranges indicated in each
    panel (including limit-cycle-induced shifts).  Only every third
    point in the data light curves is shown for clarity.  The top
    panel shows absorption due to \htwo\ and H; the bottom panel is
    absorption by \htwo.  The large gaps are caused by scan platform
    slews as described in the text.}
    \label{fig:vonesoling}
\end{figure}

Figure~\ref{fig:vonesoling} shows the \htwo\ and H profiles retrieved
from the Voyager~1 solar ingress occultation.  The multiple slews of
the scan platform are the cause of the large gaps in the profiles.
The difficulty in relating the reference spectra to the attenuated
spectra across these slews likely leaves some residual artifacts in
the profiles; nonetheless, they compare well with profiles from the
other Voyager UVS occultations.  As the only profiles measured in the
polar region of Saturn, we consider the extra work required to
determine these profiles worth the effort.  However, because only the
\htwo\ and H profiles can be retrieved, we do not carry out any
photochemical modeling of this occultation.

\clearpage

\subsection{Further Considerations From the Data Retrievals}
\label{sec:further}

At this point, it is illustrative to compare the retrievals from all
six occultations.  However, to do this we have to place the
occultation results on a common scale.

\begin{figure}[p]
    \includegraphics[width=6.5in]{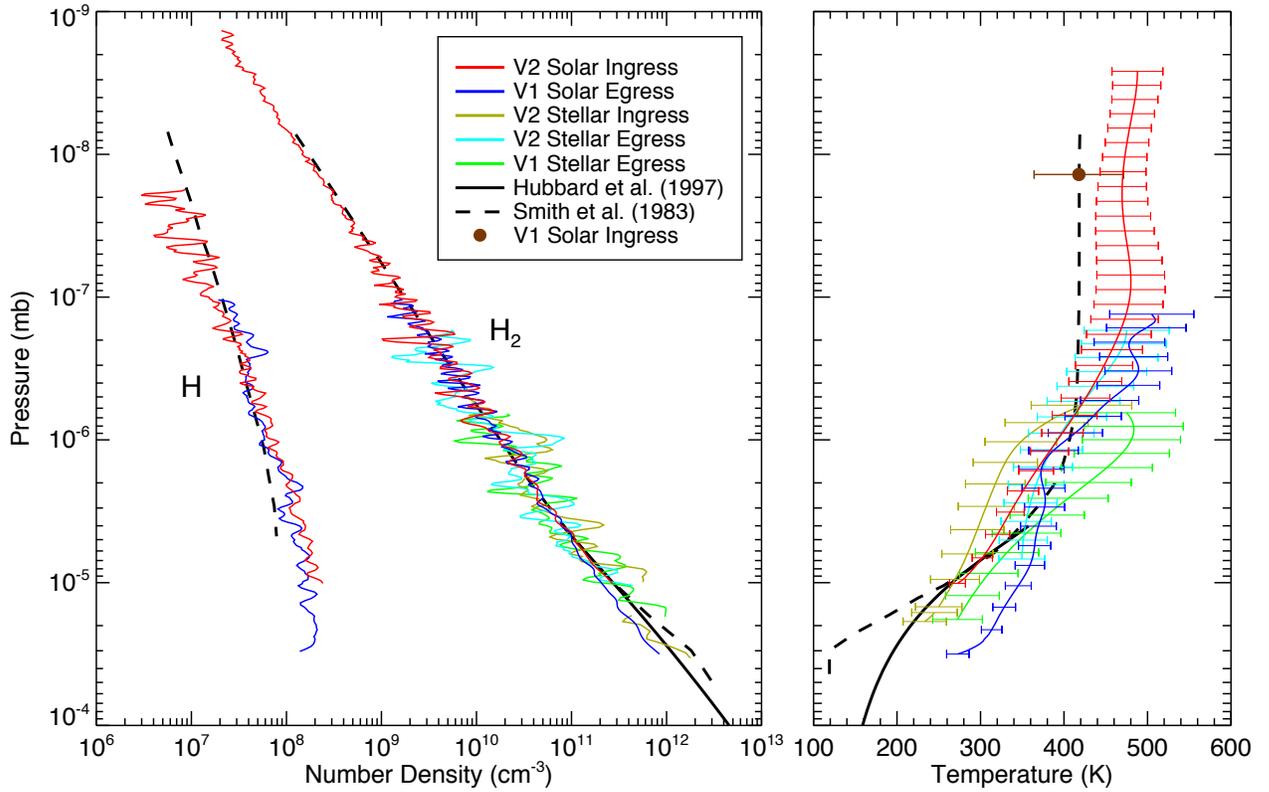}
    \caption{Summary figures showing the retrieved \htwo\ and H
    density profiles (left panel) and inferred temperature profiles
    (right panel) as a function of pressure.  Comparisons are made to
    the analyses of \protect\citet{Smith83} and
    \protect\citet{Hubbard97}.  The pressure and
    temperature derivations, as well as the implications of the
    comparisons, are discussed in the text.}
    \label{fig:presscaled}
\end{figure}

One such common scale is pressure, and in Fig.~\ref{fig:presscaled} we
have plotted the retrieved \htwo\ profiles for five of the
occultations versus pressure.  To derive the pressure, the retrieved
profiles are smoothed to eliminate the point-to-point noise and then
integrated according to
\begin{equation}
  P(r) = \int_r^\infty m_{\rm H_2} g(r') n(r')\, dr' {\rm ,} \label{eq:pres}
\end{equation}
where $m_{\rm H_2}$ is the mass of molecular hydrogen (dominates the mass
profile over our retrieved range) and $g(r)$ is gravity.  The density
profiles are assumed to fall off exponentially at the top. Because of
the large gaps in the Voyager~1 solar ingress occultation, we cannot
derive a pressure profile for this occultation.

\clearpage

Fig.~\ref{fig:presscaled} shows that retrieved \htwo\ density profiles
for the five occultations compare well, as do the H profiles for the
two solar occultations.  Also shown on the figure are the profiles
from the \citet{Smith83} and \citet{Hubbard97} analyses.  In general,
the agreement with the \citeauthor{Smith83} results is reasonably
good, although there are deviations near the top and bottom of the H
profile and at the bottom of the \htwo\ profile.  The latter
difference is notable because this is the region where
\citeauthor{Hubbard97} had difficulty matching the
\citeauthor{Smith83} results.  Indeed, there is a clear deviation of
the \citeauthor{Smith83} \htwo\ profile from the
\citeauthor{Hubbard97} profile.  Our retrieved \htwo\ profiles, as
well as the \citeauthor{Smith83} profile, merge well with the
\citeauthor{Hubbard97} profile at pressures less than 10 nanobars.
This consistency at high altitudes but not low altitudes suggests that
perhaps the \citeauthor{Smith83} retrieval pushed the transmission
spectra a bit too far at the bottom end of the range.  We restrict our
retrievals to transmission levels between 0.1 and 0.9 because outside
this range, the uncertainties in the retrievals rapidly increase as
the absorption signatures approach the signal-to-noise limits of the
transmission spectra.  \citeauthor{Smith83} presented model profiles
rather than retrieved profiles, so there were no uncertainties
presented, and we cannot say for sure if this is the case.  In any
event, the good agreement between the retrievals presented here and
the \citeauthor{Hubbard97} profile in the region of overlap, as well
as between the \citeauthor{Smith83} and \citeauthor{Hubbard97}
profiles away from pressures greater than 10 nanobars, demonstrates
that the UVS and ground-based occultation datasets are compatible and
can merge smoothly.

An advantage of determining the pressure at each radius is that we can
also get an estimate of the temperature profile from the density
profile using the ideal-gas-law relationship $P=nkT$.  In this case,
we assume the total density $n$ is just the \htwo\ density, a
reasonable assumption for the range we are considering, and solve for
$T$ at each pressure.  Uncertainties in the temperature can be
determined by propagating the uncertainties in the density through the
entire process of integrating to get $P$ and then solving for $T$.

The resultant temperature profiles are shown in right panel of
Fig.~\ref{fig:presscaled}.  It should be noted that these are not
exact representations of the temperature profile but are rather
general indications.  The feedback between the assumed temperature
profile and the \htwo\ band absorption cross sections in the retrieval
process makes it difficult to pin the temperature down in the
\htwo\ band absorption region with great accuracy.  High-resolution
measurements that reveal the detailed ro-vibrational band structure of
\htwo\ are required and must be combined with modeling of the sort
described in \citet{Hallett05rot}.  Such high-resolution spectra are
beyond the capability of the Voyager UVS and Cassini UVIS instruments,
although the UVIS does provide higher resolution than the UVS.
Nevertheless, the general comparison among the various profiles
indicates on a basic level the potential variation in the temperatures
at these pressure levels.

All of the UVS temperature profiles exhibit a generally similar shape
in the region of overlap, suggesting that latitudinal variations in
temperature at these levels of the atmosphere are relatively small
($\lta$ 100~K) but still exist.  Some wave-like structures are
suggested in both the density retrievals and the inferred
temperatures; however, these small-scale ``wiggles'' should be taken
with a grain of salt given that the temperature retrievals in
particular are only bulk estimates, as described above.

Also shown in Fig.~\ref{fig:presscaled} are the temperature profiles
of \citet{Smith83} and \citet{Hubbard97}. In the region of overlap,
the agreement between the temperature profiles inferred from our work
and that of \citet{Hubbard97} is good, although the Voyager~1 solar
egress profile does appear to show more deviation at lower altitudes
than the other profiles do (note that the same is true at the lower
altitudes of the density-pressure profile on the left).  The
\citet{Smith83} temperatures agree well in the middle of the profile,
but their profile deviates at the lowest and highest altitudes.  We
have already addressed the deviation of the \citeauthor{Smith83}
density-pressure profile from the \citet{Hubbard97} profile, so it is
not surprising to find similar deviations in the temperatures at low
altitudes.  Again, we do not know the exact circumstances by which
\citet{Smith83} derived their temperature profile, but there does not
appear to be supporting evidence for their low-altitude structure in
the Voyager UVS datasets, which on the whole are more consistent with
the \citet{Hubbard97} profile.  As for the differences between our
results and the \citet{Smith83} profile at higher altitudes, we find
slightly higher temperatures: roughly 460-490~K versus the 420~K
determined by \citet{Smith83}.

Our results can also be compared to the thermospheric temperatures
derived from the Cassini UVIS solar and stellar occultations
\citep{Shemansky12,Koskinen13} and to temperatures inferred from
auroral H$_3^+$ emission
\citep[e.g.,][]{Melin07,Melin11,Stallard12,Odonoghue14}.  The
temperatures at the highest altitudes of the Voyager~2 solar ingress
occultation are derived from the \htwo\ densities determined from the
\htwo\ continuum region.  Because the \htwo\ continuum cross sections
are generally insensitive to temperature, the density determination
and subsequent temperature inferral for the Voyager~2 solar ingress
occultation are robust results.  \citet{Koskinen13} performed a
similar analysis of the Cassini UVIS solar occultations in the
\htwo\ continuum region, and they derived exospheric temperatures
ranging from 370 to 540 K, which appear to be fully consistent with
our results.  Koskinen et al.~also find a statistically significant
difference between exospheric temperatures at high and low latitudes,
with the poles being 100-150 K warmer than the equator.  Our Voyager 2
solar ingress and Voyager 1 solar egress results are consistent with
results from similar latitudes in the \citet{Koskinen13} analysis, but
our derived temperature for the $-83\deg$ Voyager 1 solar ingress is
closer to 400 K than the $\sim$540 K temperature derived for a similar
latitude by \citet{Koskinen13}.  The single thermospheric temperature
value for this Voyager~1 solar ingress occultation is derived through
fitting an exponential profile to the \htwo\ densities in the radial
range 56900--57300~km, where the best-determined \htwo\ densities are
located for that occultation.  The temperature is then obtained from
the scale height of the fit, and the pressure for plotting purposes is
inferred from $P=nkT$.  Given the spacescraft slewing and potential
anomalous channel behavior during the Voyager 1 solar egress, we do
not place much significance on this high-latitude discrepancy between
our results and those of \citet{Koskinen13}.

\citet{Shemansky12}, in contrast, have analyzed Cassini UVIS stellar
occulations, for which signal in the \htwo\ continuum region is
unavailable owing to absorption of the stellar source by
interplanetary hydrogen.  The UVIS temperature determination at the
higher altitudes is therefore based on complex modeling of \htwo\ band
absorption \citep[e.g.,][]{Hallett05rot,Shemansky12}.  Deriving
temperatures from such modeling is difficult without high
spectral-resolution data, and --- as we noted in
section~\ref{sec:vtwosoling} --- the temperature can vary over a great
range and have only small effects on the \htwo\ densities.  From an
anaysis of the occultation spectra over the full available wavelength
range, including the \htwo\ band region, \citet{Shemansky12} derive
thermospheric temperatures ranging from 318 to 612 K for three stellar
occultations at latitudes of $-42.7\deg$, $-3.6\deg$, and $15.2\deg$,
with the near-equatorial results being the hottest and the higher
latitude results being the coldest, in contrast to the general trend
found by \citet{Koskinen13}.  Given the small number of occultations
analyzed in both our Voyager analysis and that of \citet{Shemansky12},
discrepancies between their results and ours may not be meaningful,
but we do note that the thermal profiles derived by
\citet{Shemansky12} exhibit more variability from occultation to
occultation than we see with our retrieved temperatures.

Temperatures have also been inferred from H$_3^+$ auroral emission
\citep[e.g.,][]{Melin07,Melin11,Stallard12,Odonoghue13,Odonoghue14},
where temporal and hemispheric variability have been noted.
Temperatures derived this way have ranged from 380$\pm$70 K in 1999
and 420$\pm$70 K in 2004 \citep{Melin07}, to 440$\pm$50 K in 2008
\citep{Melin11}, to 560--620 K in 2007, with the latter observations
showing significant variability on time scales of a few hours
\citep{Stallard12}.  Recent observations from 2011 reported by
\citet{Odonoghue14} suggest that the southern aurora at 583$\pm$13 K
is on average hotter than the 527$\pm$18 K northern aurora, perhaps
because the thermospheric heating rate from Joule heating and ion drag
is inversely proportional to magnetic field strength, and the larger
field strength in the north results in less total heating.  The
Voyager UVS data and the Cassini UVIS data reported to date are too
sparsely sampled to comment on whether there is a statistically
significant difference bewteen the derived tempertures in the north
versus south.

\begin{figure}[p]
    \includegraphics[width=6.5in]{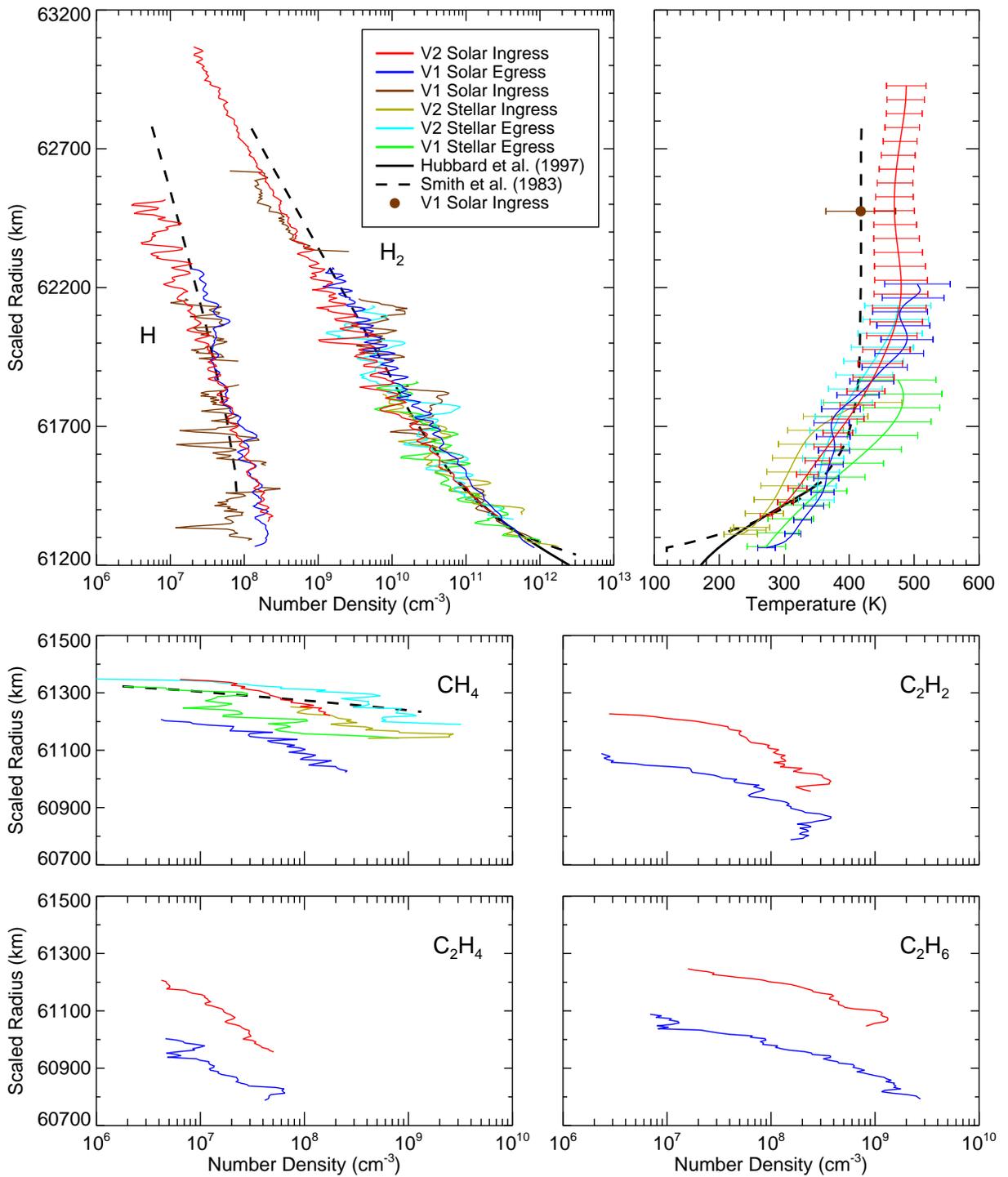}
    \caption{Summary figures showing the retrieved \htwo\ and H
    density profiles (top left panel), hydrocarbon profiles (bottom
    four panels), and inferred temperature profiles (top right panel)
    on an ``equatorial equivalent'' radial grid.  Comparisons are made
    to the analyses of \protect\citet{Smith83} and
    \protect\citet{Hubbard97}.  Derivation of the radial grid and
    implications of the comparisons are discussed in the text.}
    \label{fig:radscaled}
\end{figure}

Because the retrieved \htwo\ profiles do not extend to the levels at
which the hydrocarbons are retrieved, we cannot plot the retrieved
hydrocarbons as a function of pressure.  Instead, we show in
Figure~\ref{fig:radscaled} the profiles retrieved from all six
occultations after they have been shifted in radius to a common
reference grid for comparison and again smoothed to minimize the the
profile-to-profile noise.  The process is carried out by shifting each
\htwo\ profile in radius only until it gives the best match to the
final model of \citet{Hubbard97} in the region of overlap between the
\htwo\ density profiles as determined by a chi-squared comparison.
The shift in radius determined for \htwo\ is then applied to the
radial scales for all the density profiles for that occultation.
Shifting the profiles in this manner establishes a crude ``equatorial
equivalent'' scale but does provide a means for comparing the density
profiles for the occultations to each other and to other data or
models referenced in a similar fashion.  The \citet{Hubbard97} profile
is chosen as the reference because it is a commonly cited profile.

\clearpage

As seen in Figure~\ref{fig:radscaled}, the \htwo\ and H profiles
compare favorably for all the occultations and merge relatively
smoothly with the \citet{Hubbard97} profile, not unexpected given the
good agreement among the profiles as a function of pressure shown in
Fig.~\ref{fig:presscaled}.  Figure~\ref{fig:radscaled} also shows the
profiles from \citet{Smith83}.  The deviation of the
\citeauthor{Smith83} \htwo\ profile from the \citeauthor{Hubbard97}
profile is again clearly seen near the bottom of the \htwo\ profiles
but is not exhibited by the \htwo\ profiles retrieved in the current
analysis.  We find that the only way we can reproduce the turn-up at
the bottom of the \citeauthor{Smith83} profile is to retrieve
\htwo\ independently of \chfour.  In other words, it is possible that
the \citeauthor{Smith83} profile for \htwo\ includes some opacity that
should have been assigned to \chfour\ and vice versa.  Thus, we
conclude that the \citeauthor{Smith83} profile is in error at the
lowest altitudes and that the Voyager UVS data are consistent with the
results of \citet{Hubbard97}.

Although the \htwo\ and H profiles compare well, the hydrocarbon
profiles are a different story.  Most surprising is the shift between
the Voyager~1 solar egress (blue) and Voyager~2 solar ingress (red)
profiles.  When the radial scales are shifted such that the \htwo\ and
H profiles agree well, there is a 150--200 km altitude discrepancy
between the hydrocarbon profiles despite the similar shapes of the
profiles.  \citet{Shemansky12} saw similar large variability in the
hydrocarbon profiles from different occultations in their Cassini UVIS
analysis.  Although problems related to the determination of the
radii, perhaps caused by errors in the pointing knowledge, cannot be
ruled out, we think it is more likely to be a dynamical effect.  As is
discussed in section~\ref{sec:vonesoleg}, a variable eddy diffusion
coefficient as a function of latitude is certainly plausible and could
explain the differences in the \chfour\ profiles for all of the
occultations.  Similarly, a large-scale circulation pattern could
cause this variation, with methane restricted to high pressures in
downwelling regions, and methane carried to low pressures in upwelling
regions.  If this process is occurring, temperatures may also be
affected through adiabatic heating and cooling, and a detailed
comparison of the CIRS limb data that allow retrievals of temperatures
and hydrocarbon abundances in the middle atmosphere
\citep[e.g.,][]{Guerlet09,Guerlet10} with the numerous Cassini UVIS
occultation data that will allow determinations of temperatures and
abundances in the mesopause region \citep{Shemansky05,Nagy09} might
prove fruitful in deriving or constraining middle-atmospheric
circulation on Saturn.  Further comparison with the Voyager data
described here would shed light on the time scales and seasonality of
that circulation.  Comparisons with stratospheric general circulation
models \citep[e.g.,][]{Friedson12} might also shed light on the
observed behavior.

Figure~\ref{fig:radscaled} also shows the temperature profiles derived
from the \htwo\ density profiles.  These are the same temperature
profiles shown in the right panel of Fig.~\ref{fig:presscaled}, just
plotted on the scaled radius grid.  It is interesting to note that the
Voyager~2 solar ingress temperature profile is shifted with respect to
the Voyager~1 solar egress profile in the lower portion of the
profiles by about the same radial amount as the shift noted for the
hydrocarbon profiles.  This similarity in the vertical shift for both
the hydrocarbons and temperatures is consistent with the cooling being
correlated with the C$_2$H$_x$ hydrocarbon abundances.  A similar
correlation between the temperature profiles and the \chfour\ profiles
for the other occultations may also be the case, though within the
uncertainties of the temperature profiles, that possibility is not
clear.

The ultimate heat source driving the high thermospheric temperatures
on the giant planets is currently unknown \citep[see the reviews
  of][]{Yelle04,Nagy09}.  As such, it is difficult to speculate about
the source of the variability seen in the retrieved temperature
profiles for the different occultations.  The heat source itself may
be temporally or spatially variable, such as an auroral or
Joule-heating source at high latitudes accompanied by dynamical
redistribution of that heat
\citep[e.g.,][]{Smith07,Smith08,MullerWodarg06,MullerWodarg12}.  Waves
propagating through the upper atmosphere could introduce structure to
the profiles, although thermal conduction would smooth out such
structures on the time scale of hours (I.~M\"uller-Wodarg, personal
communication, 2008).  The major coolants H$_{\rm 3}^{\rm +}$ in the
thermosphere and \ctwohtwo\ (and \chfour, \ctwohsix) in the mesopause
region have abundances that vary with location and time.  Given that
the methane homopause level appears to vary across the planet (see
Fig.~\ref{fig:radscaled}) and that the hydrocarbons are effective
coolants at these altitudes, one might expect the thermosphere to
begin at lower altitudes (higher pressures) in regions where the
methane homopause is lowest.  There is evidence for this supposition
in the retrieved profiles shown in Fig.\ref{fig:radscaled}, but other
physics is obviously playing a role as well.  Overall, the variability
in the thermal structure with location and/or time is not very
dramatic and appears to be the most prominent at lower-thermospheric
altitudes.

\section{Concluding Remarks}
\label{sec:remarks}

Both the density retrievals and the forward models suggest
considerable compositional and some temperature variability in the
homopause region on Saturn as a function of time and/or location.  The
canonical view of vigorous atmospheric mixing in Saturn's middle
atmosphere does not hold true for all occultations at all latitudes.
The inferred pressure levels for the half-light points for the light
curves at methane-sensitive wavelengths differ by as much as a factor
of 70 between the Voyager occultations.  These differences likely
result from atmospheric dynamics.  Vertical transport via either
turbulent diffusion or vertical winds seems to be variable with
latitude/time on Saturn, significantly altering the vertical profile
of methane and the other hydrocarbons at different latitudes/times.
Results from the analysis of the first Cassini UVIS stellar
occultations (\citeauthor{Shemansky12}, \citeyear{Shemansky12}; see
also \citeauthor{Nagy09}, \citeyear{Nagy09}) support this view, with
the inferred methane homopause level being particularly deep in the
atmosphere at $-42.7\deg$ latitude at the time of the UVIS $\delta$
Ori occultation.

Ultraviolet occultations provide critical atmospheric structural
information in an atmospheric region difficult to probe by any other
means.  Our analysis suggests that UV occultations might be a
particularly sensitive probe of vertical winds and/or atmospheric
mixing at high altitudes.  The six Voyager UVS occultations do not
represent a large enough sample to provide constraints on middle
atmospheric circulation on Saturn or even to make general observations
with regard to latitude or seasonal trends in the methane homopause
level.  The two northern hemisphere occultations appear to show
methane carried to higher altitudes (lower pressures) than the
southern hemisphere occultations, but this observation is not
statistically significant.  There appears to be no correlation with
time of day, despite the fact that the two occultations with the
highest inferred methane homopause altitudes (i.e., Voyager~2 solar
ingress and Voyager~2 stellar egress) were acquired at local dusk and
slightly before midnight.  Hydrocarbon chemical lifetimes are longer
than a Saturn day throughout the atmosphere, so there should be no
diurnal variations due to chemistry.  If atmospheric tides or diurnal
winds are causing the methane concentration variability, very large
vertical winds with $\sim$10 meter per second diurnal variation would
be required to explain the observations of the Voyager~2 solar ingress
(dusk) compared with the Voyager~1 solar egress (dawn).  Such strong
winds are not predicted by the general circulation models
\citep{MullerWodarg06,MullerWodarg12,Smith07,Smith08,Friedson12}, nor
are they seen in the Earth's mesosphere from atmospheric tides.  The
most likely culprit of the observed variability is meridional
circulation in the middle atmosphere, with more modest vertical winds
operating over longer time scales
\citep[e.g.,][]{Conrath90,Barnet92,Guerlet09,Guerlet10,Friedson12,Sinclair13,Sinclair14}.
The homopause level would be suppressed in downwelling regions and
inflated in upwelling regions.  Vertically propagating waves could
also affect the profiles, and evidence for such waves in the middle
atmosphere is provided by ground-based and Cassini mid-infrared
observations \citep[e.g.,][]{Fouchet08,Orton08,Guerlet11,Li11}.

The Cassini UVIS experiment has obtained numerous solar and stellar
occultations for Saturn, and results are starting to appear in the
literature \citep{Nagy09,Shemansky12,Koskinen13}.  The full UVIS data
set promises to reveal much more about the structure and variability
of the homopause region on Saturn.  Used in combination with
temperature structure information and maps acquired from the Cassini
radio occultations and CIRS observations, middle-atmospheric
circulation patterns may be revealed for the first time on Saturn.
Comparisons with our Voyager-era analysis may shed light on
time-variable processes in Saturn's upper atmosphere.

%% Using an acknowledgements command is not in the Elsevier template,
%% but it can be used.
\ack
We thank Bill Hubbard for providing the refractivity profiles from the
ground-based observations of the 28 Sgr occultation by Saturn, as well
as providing his calculations of the 1-bar radius as a function of
latitude, Don Shemansky for providing the preliminary results from his
analysis of the $-43\deg$ latitude Cassini UVIS occultation and for
helpful discussions, Tommi Koskinen and Jacques Gustin for useful
conversations, and Chris Brion, Glyn Cooper, Peter Smith, and John
Samson for providing their highest resolution photoabsorption cross
section data.  This work has benefitted from collaborative discussions
during the ISSI workshop ``ISSI International Team on Saturn Aeronomy,
Number 166''.  Support for this project was provided by NASA Planetary
Atmospheres grant NAG5-11051 to The Johns Hopkins University Applied
Physics Laboratory and NASA Outer Planets Research Program grant
NNG05GG65G to the Lunar and Planetary Institute (LPI).  J.~Moses also
acknowledges support from LPI during the initial stages of this
project (LPI is operated by the Universities Space Research
Association under contract with NASA), and for recent support from
NASA OPR grant grant NNX13AK93G.

\section*{Appendix A. Sensitivity Tests: How Unique Are Forward Models?}
\label{sec:appendix}

Most of the ultraviolet occultation analyses for Saturn to date have
used some kind of forward-modeling technique
\citep{Broadfoot81sat,Sandel82sat,Festou82,Smith83,Shemansky12}, as
well as other, sometimes simplified, equations and arguments to back
out the atmospheric scale height, temperatures, and densities in the
occultation regions.  Exceptions are the retrievals from
\citet{Koskinen13} and our own work.  The high sensitivity of the
forward models to uncertain model inputs --- along with noisy data ---
may explain some of the differences between the derived exospheric
temperature and density structure from these various groups (see
section~\ref{sec:stellardata}).  As is discussed in
section~\ref{sec:forward}, the forward models are nonunique, and
assumptions about the temperature variation with altitude, the
atmospheric mean molecular mass and its variation with altitude, the
zonal wind velocities and their variation with altitude and latitude
(and other factors that influence the gravitational acceleration), and
the 1-bar (or other constant pressure) radius and its variation with
latitude can strongly affect the inferred properties of the
occultation regions.  The first three parameters --- temperature, mean
molecular mass, gravitational acceleration --- all affect the
atmospheric scale height and control how extended or compressed the
atmosphere is.  The last parameter (radius of geopotential surface)
affects not only the assumptions about the gravity profile but also
helps set the scale for the occultation observations, from which
radial profiles are obtained.  In this appendix, we illustrate the
sensitivity of the models to certain changes in the model assumptions.

One of the main uncertainties in the models is the temperature
structure in the lower atmosphere at altitudes below (or above) those
in which the occultations are most sensitive.  When constructing a
forward model, investigators generally fix their base ``reference''
level at some region in which the temperature and radius are
reasonably well known for some pressure level.  This reference-level
information is needed to set the boundary conditions for solving the
hydrostatic equilibrium equation.  For the giant planets, the 1-bar or
100-mbar pressure levels, the ``cloud tops'', or some level defined by
the light-curve transmission have all been used for this reference
level.  The density-radius structure away from this base level can be
derived using the hydrostatic equilibrium equation, after making
assumptions about the thermal structure and other model parameters in
the intervening region.  In our forward models, as in those of
\citet{Shemansky12}, we use the 1-bar radius as our base level, and we
build the model by making assumptions about the temperature profile
from 1 bar on up to the occultation region.  This procedure is
particularly dangerous, as it relies on a good knowledge of all the
atmospheric parameters from the 1-bar region over many scale heights
up to the occultation altitudes.

Measurements of the temperature profile in Saturn's troposphere and
stratosphere from Voyager RSS or IRIS observations were not obtained
at the same time and location as the UVS occultations, although some
overlap in the latitude coverage of the Voyager IRIS data and the UVS
occultation regions does exist
\citep[e.g.,][]{Conrath83,Conrath84,Conrath90,Conrath98}.  Moreover,
temperature derivations from RSS and IRIS data do not extend to
pressures less than $\sim$0.5 mbar, leaving gaps in our knowledge of
the thermal structure.  Ground-based stellar occultation observations
\citep[e.g.,][]{Hubbard97} or high-spectral-resolution infrared
observations can potentially fill these gaps, but no such observations
were performed at the time of the Voyager UVS occultations.
Therefore, our knowledge of the thermal structure at the occultation
latitudes is sparse, and the forward models that use the 1-bar level
or cloud tops as their base level are poorly constrained.

This situation is improved in the Cassini era, as numerous
occultations have been recorded by Cassini UVIS
\citep[e.g.,][]{Nagy09,Shemansky12,Koskinen13}, and several other
techniques and instruments are available to help constrain
temperatures and atmospheric structure at the occultation latitudes.
Of particular note are the CIRS limb observations
\citep{Fouchet08,Guerlet09,Guerlet10,Guerlet11}, which can reliably
track stratospheric temperatures up to $\sim$10$^{-2}$ mbar, and the
VIMS stellar occultations
\citep[e.g.,][]{Nicholson06,Bellucci09,Kim12}, which should eventually
help place stratospheric radius scales on a pressure grid.

\begin{figure}[p]
    \includegraphics[scale=0.6,angle=-90]{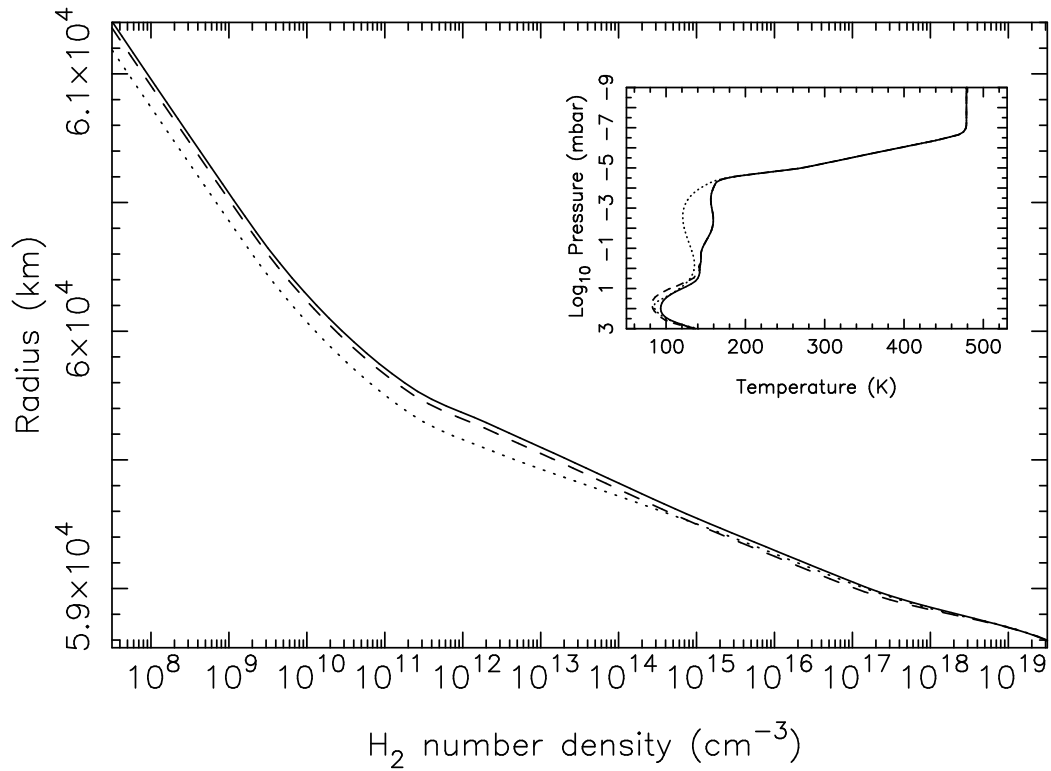}
    \caption{The H$_2$ density profile derived from solving the
    hydrostatic equilibrium equation for three different assumptions 
    about the temperature profile (insert).  All other model parameters
    are held constant.  Differences in the assumed temperature structure 
    in the lower atmosphere can greatly affect densities at the higher 
    altitudes probed by the ultraviolet occultations.}
    \label{fig:appone}
\end{figure}

For the Voyager occultations, the lack of constraints on the
tropospheric and stratospheric temperatures can complicate the
derivation of temperatures in the occultation regions from forward
models.  For example, Figure~\ref{fig:appone} shows three fictitious
temperature profiles, and the resulting density structure when all
other atmospheric parameters (e.g., rotation rate, mean molecular
mass, latitude, 1-bar radius) are held constant.  The model
represented by a solid line assumes the thermal structure derived by
the Cassini CIRS limb observations for $-30\deg$ planetographic
latitude \citep{Guerlet09} from 1 bar to 6$\scinot-5.$ mbar, blending
to an ad hoc thermospheric profile at pressures less than
6$\scinot-5.$ mbar.  The model represented by a dashed line uses the
colder Voyager radio occultation profile as reported by
\citet{Lindal92} for pressures greater than 0.5 mbar, but is otherwise
identical to the model represented by a solid line.  The model
represented by a dotted line assumes a preliminary Cassini UVIS
temperature profile from $-42.7\deg$ latitude as reported by
\citet{Shemansky05}, which has since been revised \citep{Shemansky12},
from 1 bar to 7$\scinot-5.$ mbar, but again blends into the ad hoc
thermospheric profile at pressures less than 7$\scinot-5.$ mbar.  Note
that the stratosphere in this model is much colder than the other two
models. All model calculations were performed for $-27\deg$
planetocentric latitude, assuming a 1-bar radius of 58770.6 km, an
altitude-independent rotation rate of 10.55 hr (which includes zonal
winds), and an altitude-independent atmospheric mean molecular mass of
2.28 amu (10\% He, 0.45\% CH$_4$).

\clearpage

The resulting \htwo\ density structure for these models is shown in
Fig.~\ref{fig:appone}.  Because of the colder stratosphere assumed for
the dotted-line model, the density at any particular radius in the
region typically probed by the UVS occultations (i.e., at
\htwo\ densities less than a few times 10$^{11}$ cm$^{-3}$) is much
smaller in this model than the other models, despite the identical
thermospheric temperature profile.  The offset in radius at any
particular density is about 25 km between the solid and dashed models
and $\sim$100 km between the solid and dotted models for
\htwo\ densities probed by the UVS occultations.  Similarly, at any
particular radius level, the \htwo\ density for the dotted model can
be as much as a factor of six smaller than that of the solid-line
model.  Assumptions about the lower-atmospheric temperature structure
can clearly affect the predicted density structure when the base level
is far from the occultation region.

Although the slope of the occultation light curves and/or retrieved
\htwo\ density profile can help the forward modelers determine the
exospheric temperature, the temperatures in the lower thermosphere are
difficult to obtain from forward models (without additional
information provided by the spectra themselves) because of
uncertainties in the temperature gradients and lower-atmospheric
temperatures.  For example, if a forward modeler were to assume the
base of the thermosphere began at higher pressures than the dotted
model shown in Fig.~\ref{fig:appone}, such that the rapid increase in
temperature with height were located deeper in the atmosphere, then
the density structure of the dotted-line model in
Fig.~\ref{fig:appone} could be brought into better agreement with the
other models.  There is clearly a trade-off between assumptions about
lower-atmospheric temperatures and assumptions about the temperature
gradient in the lower thermosphere actually probed by the ultraviolet
occultations --- if the tropospheric and stratospheric temperatures
adopted by the forward modelers are not realistic, the resulting
inferences about the lower-thermospheric temperatures will likely be
in error.  Similarly, the temperature gradients in the lower
thermosphere are not uniquely constrained from forward models.

Such non-uniqueness problems severely curtail the usefulness of
forward-modeling techniques as a means for determining temperatures in
the lower regions of the thermosphere; the hybrid retrieval technique
described in this paper is a more reliable method.  However, if the
spectral resolution of the observations were higher than that of the
UVS instrument, it might be possible to exploit temperature-dependent
spectral behavior to help derive temperatures through forward modeling
(see the Cassini UVIS analysis and discussion of \citet{Shemansky12}),
but without very high resolution that resolves the ro-vibrational
structure in the \htwo\ lines, elements of non-uniqueness would likely
remain.

\begin{figure}[p]
    \includegraphics[scale=0.6,angle=-90]{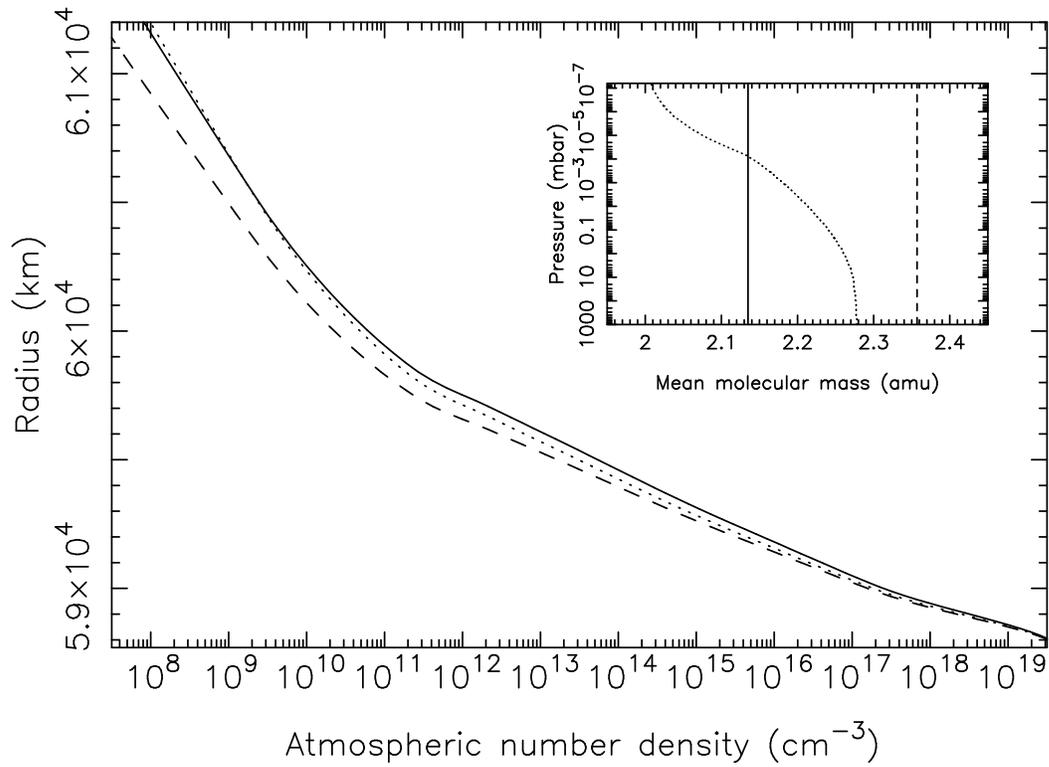}
    \caption{The atmospheric density profile derived solving the
    hydrostatic equilibrium equation for three different assumptions 
    about the atmospheric mean molecular mass profile (insert).  All 
    other model parameters are held constant.  Differences in the 
    assumed atmospheric mean molecular mass and its variation with 
    altitude can greatly affect densities at the higher 
    altitudes probed by the ultraviolet occultations.}
    \label{fig:apptwo}
    \label{lastfig}
\end{figure}

The temperature structure is not the only model parameter that can
have a large effect on the resulting density structure.  The mean
molecular mass also affects the atmospheric scale height, and given
the uncertainties in the helium abundance and its variation with
altitude on Saturn, the resulting uncertainties in the atmospheric
density structure are large.  Models with three different assumed
mean-molecular-mass profiles are shown in Fig.~\ref{fig:apptwo}.  The
models have the same temperature structure and other model parameters
(see above) as shown by the solid curve in Fig.~\ref{fig:appone}.  For
the model represented by a solid line in Fig.~\ref{fig:apptwo}, we
assume that the atmospheric mean molecular mass is constant with
height, with a value 2.135 amu (i.e., 6\% He, 94\% H$_2$, no methane),
as was assumed by \citet{Hubbard97} and several other Voyager-era
investigators.  For the model represented by a dashed line in
Fig.~\ref{fig:apptwo}, we assume a constant value of 2.357 amu (i.e.,
14\% He, 0.45\% CH$_4$, with the rest being H$_2$), the upper limit
from \citet{Conrath00}.  The dotted line represents a more realistic
model in which we assume a mean molecular mass of 2.278 amu (10\% He
and 0.45\% CH$_4$ in the troposphere) but use a
photochemistry/diffusion model to calculate the fall off in mean
molecular mass with altitude due to molecular diffusion.
Fig.~\ref{fig:apptwo} shows that the mean molecular mass can strongly
influence the density structure in the models.  For a constant
atmospheric density of 10$^8$ cm$^{-3}$, the difference between the
dotted model and the dashed model is a very large 275 km in radius.
For a constant radius of 61500 km, the difference in the predicted
atmospheric density between the 2.357-amu model and the others is a
factor of 2.5.

\clearpage

An incorrect assumption about the atmospheric mean molecular mass
profile can therefore lead to incorrect derivations of atmospheric
densities and temperatures.  An overestimate in the mean molecular
mass will lead to an overestimate in the temperatures needed to derive
the same H$_2$ density and light-curve absorption.  Note that both
\citet{Festou82} and \citet{Smith83} appear to assume a pure H$_2$
atmosphere in the thermosphere, which is a reasonable assumption for
this region.  However, if the actual slope of the mean molecular mass
profile varies with altitude in Saturn's thermosphere, which is
likely, this assumption could have an effect on the resulting derived
slope for the atmospheric temperatures in this region.  The assumed
mean molecular mass profile will have a larger effect on models whose
reference base level is far from the occultation region.

The local gravity is a third parameter that can affect the density
structure in forward models \citep[see][]{Koskinen13}.  Although
the radial gravity structure is more straightforward as an input
parameter in the forward models, some uncertainties do exist because
of uncertainties in zonal wind speeds and their variation with height.
Saturn's rapid rotation must be included in the terms for calculating
the gravitational acceleration at any radius \citep{Lindal85}, and the
strong zonal winds affect the local rotation rate, especially at low
latitudes on Saturn.  Adopting incorrect zonal winds speeds would
result in only a few km error in the atmospheric density profiles in
the occultation regions provided that the base level radius is
accurately constrained; however, neglecting rotation entirely in the
gravity equation can lead to much larger errors, and the winds
themselves strongly perturb the isobaric radius surfaces.  This latter
consideration is very important.  As an example, the standard
reference ellipsoid for Saturn assumes an equatorial radius of 60268
km and a polar radius of 54364 km at the 1-bar level.  At $-45.3\deg$
planetocentric latitude, that would correspond to a 1-bar radius of
57057.6 km for the ellipsoid model, as compared with the 56947.6 km
1-bar radius derived from the Cassini radio occultation data (Paul
Schinder, personal communication, 2009) --- a difference of 110 km.
By the same token, the 1-bar radius can change by ~100 km within a
one-degree latitude range at mid latitudes, which complicates forward
modeling of the occultations that span a wider range of latitudes.

The overall uncertainties in the geopotential radii for Saturn can
lead to relatively large uncertainties in temperature derivations from
the forward models, as an offset in the radius of the base level would
require a relatively large temperature adjustment to bring the H$_2$
densities in line with what is required from the light-curve
observations.  As an example, the switch from the reference ellipsoid
model to a more realistic 1-bar radius as defined by the Cassini
radio-science experiments, changes the temperatures derived for the
Cassini UVIS $-42.7\deg$ latitude occultation from 121 K to $\sim$179
K at a radius of 58149 km (i.e., a 48\% increase; D. Shemansky,
personal communication, 2009).

The bottom line from the above discussion is that forward models
cannot uniquely constrain temperatures from ultraviolet occultations
in general, not unless there is some temperature-sensitive spectral
behavior that can be exploited from the observations, and not unless
the inputs to the models are better constrained.  These limitations
must always be kept in mind when considering temperatures derived from
forward models.

\label{lastpage}

% Bibliographic references with the natbib package:
% Parenthetical: \citep{Bai92} produces (Bailyn 1992).
% Textual: \citet{Bai95} produces Bailyn et al. (1995).
% An affix and part of a reference:
%   \citep[e.g.][Ch. 2]{Bar76}
%   produces (e.g. Barnes et al. 1976, Ch. 2).-

\bibliography{references.bib}

%% Use the plainnat style for ``Icarus'' mode to display DOI numbers
%% among other things.  However, revert to the Elsevier elsart-harv
%% mode for ``Elsevier'' mode.
%\bibliographystyle{plainnat}
\bibliographystyle{elsart-harv}
%\bibliographystyle{icarus}

%% --Figures-- %%

\clearpage
%\input{figures}

%% --Tables-- 

%\clearpage
%\input{tables}

\end{document}